\documentclass[aps,prx,twocolumn,superscriptaddress,nofootinbib]{revtex4-2}

\usepackage[dvipsnames]{xcolor}
\usepackage{hyperref}
\hypersetup{
  breaklinks=true,
  colorlinks=true,
  allcolors=BlueViolet,
}

\usepackage{graphicx} 
\usepackage{svg}

\usepackage{amsmath}
\usepackage{physics,braket,bm}

\usepackage{xfrac}
\usepackage{enumerate}

\let\log\relax
\newcommand{\log}{\mathrm{log}}

\newcommand{\F}{\mathcal{F}}

\newcommand \be{\begin{equation}}
\newcommand \ee{\end{equation}}

\newcommand \bea{\begin{eqnarray}}
\newcommand \eea{\end{eqnarray}}

\usepackage{amssymb}

\usepackage[inline]{enumitem}
\setlist[enumerate]{leftmargin=*}

\newcommand {\rsub}[1]{\textcolor{black}{#1}}

\newcommand {\rsubb}[1]{\textcolor{black}{#1}}

\newcommand{\InfleqtionC}{Infleqtion,  Chicago, IL, 60615}
\newcommand{\InfleqtionM}{Infleqtion,  Madison, WI, 53703}
\newcommand{\InfleqtionB}{Infleqtion, Boulder, CO, 80301}
\newcommand{\UWM}{Department of Physics, University of Wisconsin-Madison,  Madison, WI, 53706}

\definecolor{mscolor}{rgb}{0,0.5,0.5}
\definecolor{mscolor}{rgb}{0,0.5,0.5}
\definecolor{cpcolor}{rgb}{0.4,0,0.8}
\definecolor{cpcolor}{rgb}{0.5,0,0.5}

\begin{document}

\title{Qubit syndrome measurements with a high fidelity Rb-Cs Rydberg gate}
\author{J. Miles}
\affiliation{\InfleqtionM}
\author{M. T. Lichtman}
\affiliation{\InfleqtionM}
\author{A. M.  Scott}
\affiliation{\InfleqtionM}
\author{J. Scott}
\affiliation{\InfleqtionM}
\author{S. A. Norrell}
\affiliation{\UWM}
\author{M. J. Bedalov}
\affiliation{\InfleqtionM}
\author{D. A. Belknap}
\affiliation{\InfleqtionM}
\author{D. C. Cole}
\affiliation{\InfleqtionB}
\author{S. Y. Eubanks}
\affiliation{\InfleqtionB}
\author{M. Gillette}
\affiliation{\InfleqtionM}
\author{P. Gokhale}
\affiliation{\InfleqtionC}
\author{J. Goldwin}
\affiliation{\InfleqtionM}
\author{M. Iliev}
\affiliation{\InfleqtionB}
\author{R. A. Jones}
\affiliation{\InfleqtionB}
\author{K. W.  Kuper}
\affiliation{\InfleqtionB}
\author{D. Mason}
\affiliation{\InfleqtionB}
\author{P. T. Mitchell}
\affiliation{\InfleqtionB}
\author{J. D. Murphree}
\affiliation{\InfleqtionM}
\author{N. A. Neff-Mallon}
\affiliation{\InfleqtionM}
\author{T. W. Noel}
\affiliation{\InfleqtionB}
\author{A. G. Radnaev}
\affiliation{\InfleqtionB}
\author{I. V. Vinogradov}
\affiliation{\InfleqtionB}
\author{M. Saffman}
\affiliation{\InfleqtionM}
\affiliation{\UWM}
\date{\today}

\begin{abstract}
We demonstrate an inter-species entangling Rydberg gate between rubidium (Rb) and cesium (Cs) atoms with fidelity 
$\mathcal F = 0.975\pm 0.002$. The two-species atom array enables in-place quantum non-demolition (QND) qubit measurements which are a key capability for quantum error correction. We demonstrate this functionality with multi-atom error syndrome measurements achieving QND measurement fidelities of 
${\mathcal F}_{\rm QND} = 0.933(12)$ and $0.865(17)$ for two- and three-qubit plaquettes, respectively.
\end{abstract}

\maketitle


{\it Introduction --} Neutral atom arrays have scaled to more qubits than any other modality \cite{Manetsch2025}, and have demonstrated single-species entangling operations with fidelities well above 99\% with four different atomic elements \cite{Evered2023,Radnaev2025,Peper2025,Muniz2025,Tsai2025}.
Incorporating measurement-based error correction of logical qubits without undesired disturbance of proximal data qubits when measuring ancillae has required  additional operations such as movement to a measurement zone \cite{Deist2022,Bluvstein2022},  or shelving or hiding \cite{Lis2023,Norcia2023,SMa2023,Graham2023b,Bluvstein2026,Chiu2025,BHu2025}. \rsub{These extra operations may slow down the logical cycle rate, or may add to the error budget of syndrome extraction.} 

An alternative solution that does not require any additional operations beyond the gates needed for mapping syndrome information from data to ancilla qubits is to use different atomic elements for data and ancillae. 
\rsub{This was first demonstrated with trapped ions \cite{Tan2015,Negnevitsky2018,Hughes2020} followed by  neutral atoms
\cite{Beterov2015,Anand2024}.  Although there are significant operational differences in working with  two species of trapped ions or neutral atoms, in both cases the feature that different elements respond to different wavelengths of light makes it possible to operate on one species with essentially no disturbance of the other, as was first demonstrated for neutral atoms with a Rb-Cs dual species array \cite{Singh2022}}.  This also allows one species to be used to detect environmental noise, thereby providing feedback which can improve the coherence of the other species \cite{Singh2023}. Interspecies neutral atom logic gates  can be  implemented with Rydberg interactions \cite{Saffman2010} which can be designed to work equally well for interspecies  as for intraspecies coupling \cite{Beterov2015, Samboy2017, Otto2020,Ireland2024}. 

\begin{figure*}[!t]
    \centering
    \includegraphics[width=.95\textwidth]{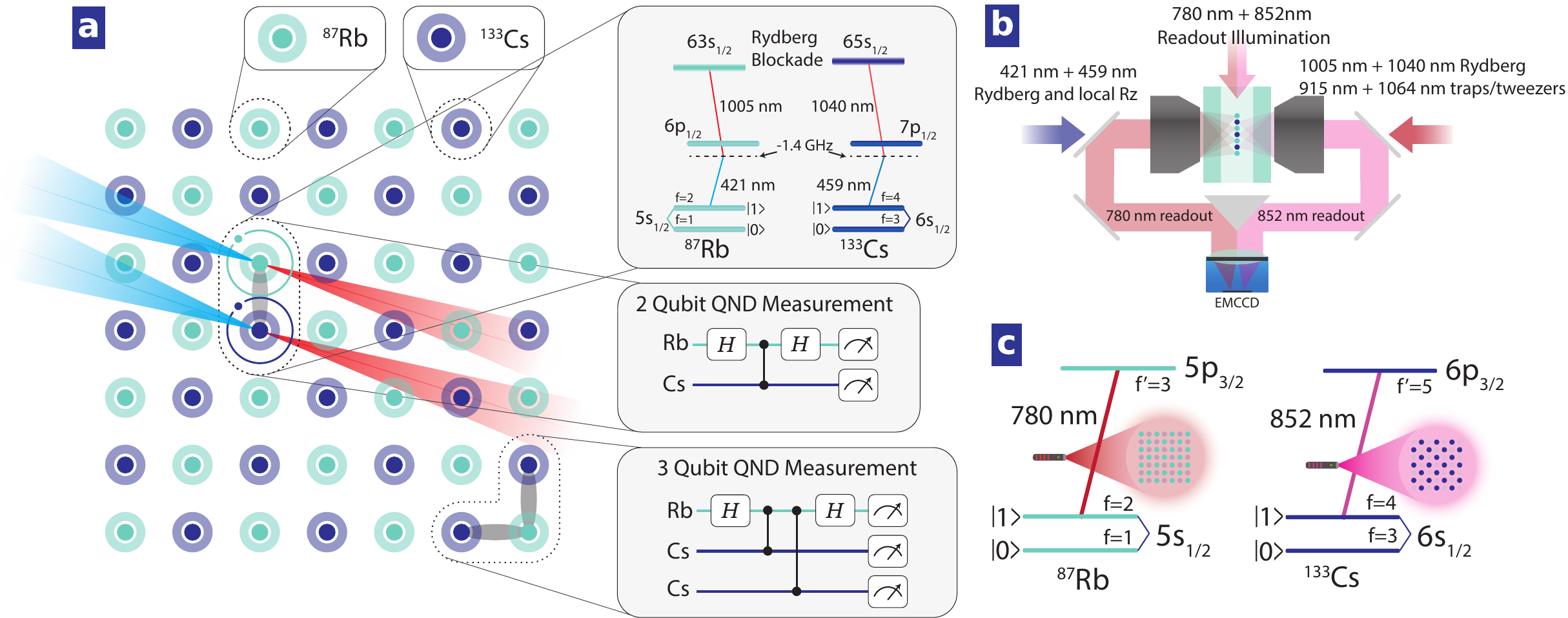}
    \caption{\textbf{a)} Dual species atom array. $^{87}$Rb and $^{133}$Cs atoms are prepared in a checkerboard pattern. Individual Rydberg state addressing beams with wavelengths 421, 1005, 459, and 1040 nm are used to excite neighboring Rb-Cs pairs to Rydberg states (top inset). Gate model circuits for performing two- and three- qubit QND measurements (middle and bottom insets). \textbf{b)} Illustration of independent species readout via two-sided imaging. Scattered photons for Rb and Cs are captured via independent optical paths and are imaged on a shared EMCCD camera. A set of dichroic mirrors and narrow-line filters are used to perform wavelength discrimination. \textbf{c)} Readout level diagrams for each species.   
    }
    \label{fig:setup}
\end{figure*}

In this Letter we demonstrate an order of magnitude improvement in the fidelity of interspecies $\sf CZ$ gates and Bell state preparation with Rb and Cs hyperfine encoded qubits in an interleaved  array of two atomic species \cite{CFang2025}. Gates are performed with tightly focused Rydberg beams that address individual atoms  \cite{Graham2022,Radnaev2025,WChung2025}. Gate fidelity of $\mathcal F = 0.975(2)$ is demonstrated \rsubb{on a single atom pair within the array} which is well above previous Rb-Cs results with hyperfine encoded qubits \cite{Anand2024}, and comparable to the recent demonstration of a Cs-Cs gate mediated by a Rb atom using ground-Rydberg encoding which achieved $\mathcal F = 0.967(17)$ \cite{White2026}. The observed fidelity is in agreement with  Monte-Carlo simulations of a detailed error model, which identifies a route to performance below the threshold for surface code error correction \cite{Miles2026SM}. 
We proceed to use the gate for a quantum non-demolition measurement of a data qubit and parity  syndrome measurements on a three atom plaquette, which is a building block for surface code error correction.

{\it Experimental apparatus --} A high-level overview of the experimental apparatus is shown in Figure \ref{fig:setup}. The vacuum system incorporates a  source region where Rb and Cs atoms are pre-cooled (Infleqtion PICAS) and are then pushed to a differentially pumped glass science cell where the atoms are cooled in a 3D magneto-optical trap (MOT) and loaded into a $7\times 7$ array of optical traps. The science cell includes internal electrodes for canceling background electric fields. The trap array of 1064 nm wavelength tweezers is produced by a spatial light modulator (SLM) that provides sites of waist ($1/e^2$ intensity radius) $w=1.7~\mu\rm m$ nominally spaced $5.85(0.02)~\mu\rm  m$ apart on a square grid. Two $\mathrm{NA}=0.7$ objective lenses provide tight focusing from two sides  for the trapping light (1064 nm),  counter-propagating Rydberg beams (421, 459, 1005, 1040 nm), atom rearrangement (915 nm) and dual-sided imaging (780 and 852 nm).

After loading the 3D MOT in the science cell, the magnetic field is zeroed and cooling lasers are further detuned  for sub-Doppler cooling in the tweezer traps. 
A single 915 nm tweezer is steered via crossed-AODs and moves both Rb and Cs into the desired pattern of occupied sites.
Qubits are encoded in hyperfine-Zeeman clock states $\ket{f,m_f}$
for $^{87}$Rb ($\ket{0}\equiv\ket{1,0}$,  $\ket{1}\equiv\ket{2,0}$) and  $^{133}$Cs ($\ket{0}\equiv\ket{3,0}$, $\ket{1}\equiv\ket{4,0}$). Both species are optically pumped to their respective $\ket{1}$ states using $\pi$ polarized D1 light (Rb: 795 nm $\ket{5s_{1/2},f=2}\rightarrow \ket{5p_{1/2},f'=2}$, Cs: 895 nm $\ket{6s_{1/2},f=4}\rightarrow \ket{6p_{1/2},f'=4}$). Both species are imaged through the objective lenses at 780 and 852 nm by focusing the optical paths to separate regions of a single EMCCD camera (N\"uV\"u HN\"{u} 512).  State-selective measurements are performed via resonant blow-away of the $\ket{1}$ states  followed by atom occupancy detection.

In order to reach sufficiently low temperatures suitable for high fidelity Rydberg gates we employed adiabatic cooling. Starting at trap depths of  $U_{\rm i}/k_{\rm B}=1.1,~2.0  ~{\rm mK} $ for Rb, Cs respectively, the initial temperature after polarization gradient cooling and optical pumping is measured by drop and recapture to be 
$T_{\rm i}= 15.4(6),~17.3(7)~\mu\rm K$ for Rb, Cs. The traps are then \rsub{adiabatically lowered \cite{Miles2026SM}} to $U_{\rm f}/k_{\rm B}=0.073,~0.136 ~\rm mK$ over a time of 1 ms using a cubic spline intensity profile. After the ramp the final atom temperature is measured with drop-and-recapture to be $T_{\rm f}= 3.2(2),~4.3(2) ~\mu\rm K$ for Rb, Cs, within approximately 20\% of the predicted final temperature.

Low atom temperatures, combined with the far-detuned 1064 nm trapping wavelength resulted in excellent $T_1$ (9.6, 33.5 s) and $T_2^*$ (44.9, 27.2 ms) coherences for Rb and Cs,  which are documented in \cite{Miles2026SM}.  
As shown there, we find no statistically significant difference between $T_2^*$ measurements taken with or without background readout of the other species. This isolation is essential for in-place syndrome measurements on decoupled registers of data and ancilla qubits, \rsub{as has been demonstrated in other dual-species arrays \cite{Anand2024,Singh2022}}. We note that the adiabatic ramp improves gate fidelity by reducing errors from atom motion but reduces gate fidelity due to the worse localization of the atoms relative to the focused Rydberg excitation beams and separation dependent variation of the Rydberg interaction strength. The supplemental material \cite{Miles2026SM} provides numerical simulations of these effects and their dependence on trap depth.

\begin{figure}[!t]
    \includegraphics[width=0.5\textwidth]{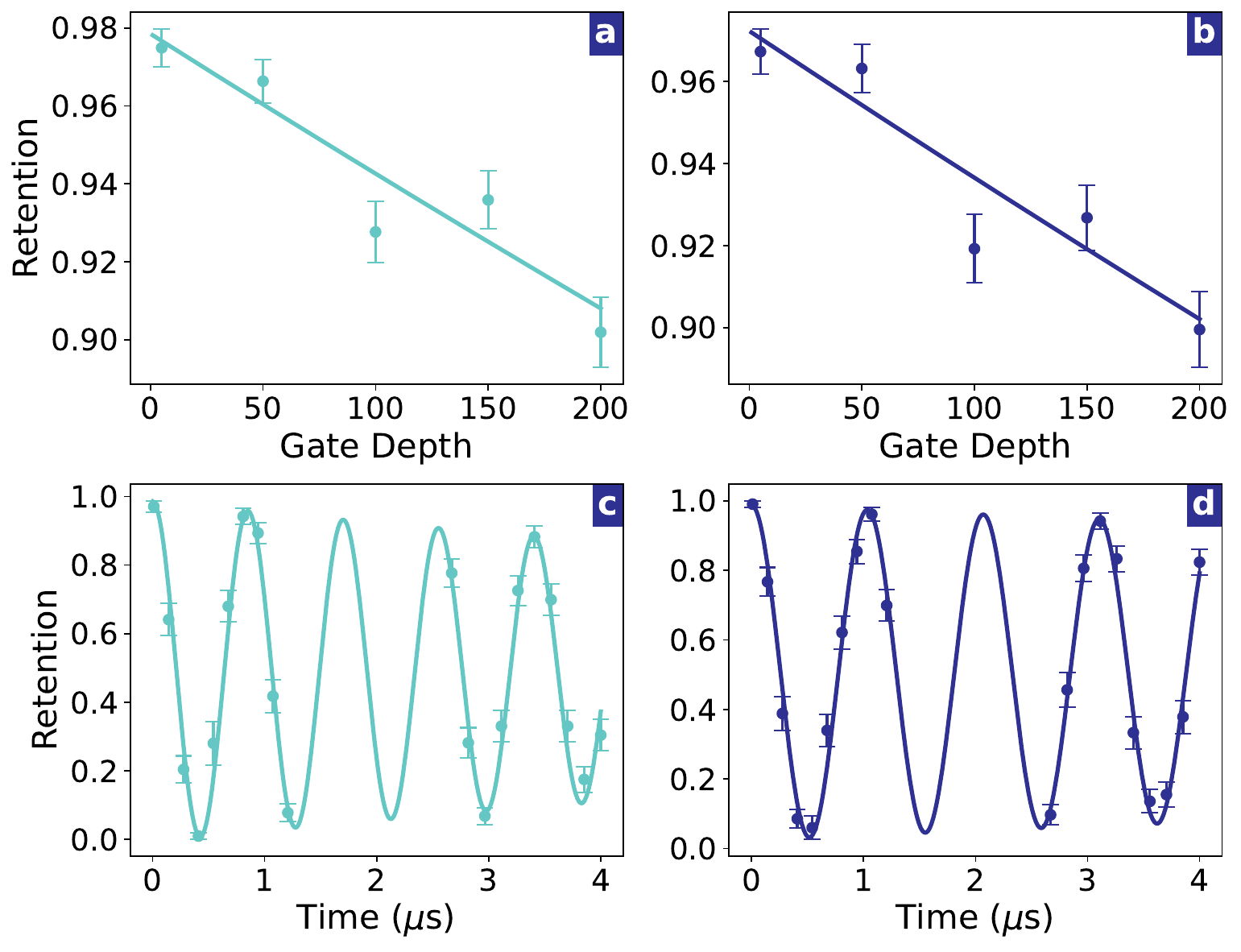}
    \caption{Characterization of single qubit operations.     
    \textbf{a)} Randomized benchmarking for one Rb qubit, yielding $\mathcal{F}=0.99963(5)$, measured simultaneously with Cs.  
    \textbf{b)} Randomized benchmarking for one Cs qubit, yielding $\mathcal{F}=0.99962(5)$, measured simultaneously with Rb.  
    \textbf{c)} Ramsey-Stark oscillations for Rb demonstrating ${\sf R_z}(\phi)$ oscillations at $f= 1.175(2)$ MHz and $f\tau = 18(3)$. See \cite{Miles2026SM} for details. 
    \textbf{d)} Same as c), but for Cs, at  $f= 0.966(2)$ MHz and $f\tau = 34(15)$.
    }
    \label{fig:rb_mw}
\end{figure}

A complete set of one-qubit gates is implemented with  the 421 and 459 nm first-photon Rydberg lasers or individual ${\sf R_z}(\phi)$ gates 
and 6.8 and 9.2 GHz microwaves for global rotation ({\sf GR}) ${\sf R}(\theta,\phi)$ gates (rotation through angle $\theta$ about an axis rotated by $\phi$ from the $x$ axis in the equatorial plane).  A single microwave horn provides global operations to both species via a high powered diplexer which combines two microwave signals, each with power 40 W, at 6.8 and 9.2 GHz that provides Rabi rates  of  $2\pi \times 9.63$ and  $2\pi \times 12.6$ kHz  for Rb and Cs qubits, corresponding to $\pi$ gate times of 51.9 and 39.6  $\mu \rm s.$ 
Randomized benchmarking over the full set of Clifford gates returned fidelities of 0.99963(5) and 0.99962(5) for Rb and Cs {\sf GR} gates, as displayed in Fig. \ref{fig:rb_mw}. Local ${\sf R_z}(\phi)$ gates were achieved via a differential light shift with the blue Rydberg laser for each species. For Rb the 421 nm laser was detuned by -1.494(2) GHz  from the $6p_{1/2}$ state while for Cs the 459~nm laser was detuned by -1.382(2) GHz  from   the $7p_{1/2}$ state.   The Ramsey data in  Fig. \ref{fig:rb_mw} show a differential light shift induced by the blue lasers of $2\pi \times1.175(2)$ and $2\pi \times0.966(2)$ MHz  affording ${\sf R_z}(\phi=\pi)$ gate times of 426 and 518 ns for Rb and Cs. 
\rsub{Here, and in all subsequent figures, error bars represent the standard deviation.}
In this work, local ${\sf R_z}$ gates were only used for state preparation in the  3-atom QND demonstration, whereas phase tracking on subsequent microwave operations \rsub{or {\sf GR} gates were} used to implement ${\sf R_z}$ gates that corrected single qubit phases in the {\sf CZ} gate.

\begin{figure}[!t]
    \includegraphics[width=0.5\textwidth]{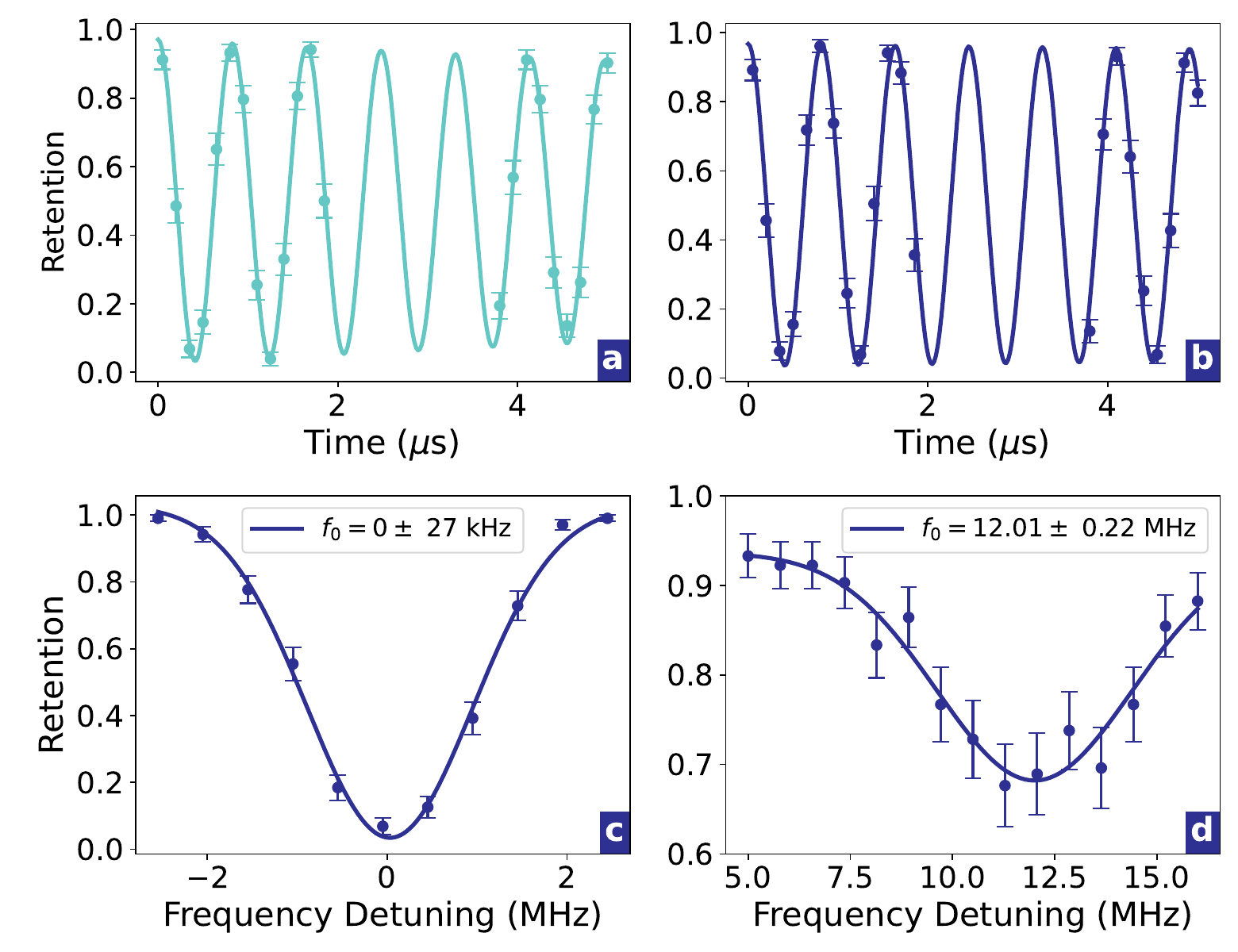}
    \caption{\textbf{a)} Rubidium ground-Rydberg Rabi oscillations with rate  $\Omega=2\pi\times1.2085(17)$ MHz.   \textbf{b)} Cesium ground-Rydberg Rabi oscillations at $\Omega=2\pi\times 1.2219(15)$ MHz. 
    \textbf{c)} Rydberg spectroscopy of Cs with a neighboring Rb atom in the ground state.     \textbf{d)} Rydberg spectroscopy of Cs with a neighboring Rb atom in the Rydberg state $\ket{63s_{1/2}, m_j=-1/2}$.  The shift in the Cs Rydberg resonance is due to the Rb-Cs interaction.}
    \label{fig:rydberg_rabi_data}
\end{figure}

\begin{figure}[!t]
       \includegraphics[width=0.45\textwidth]{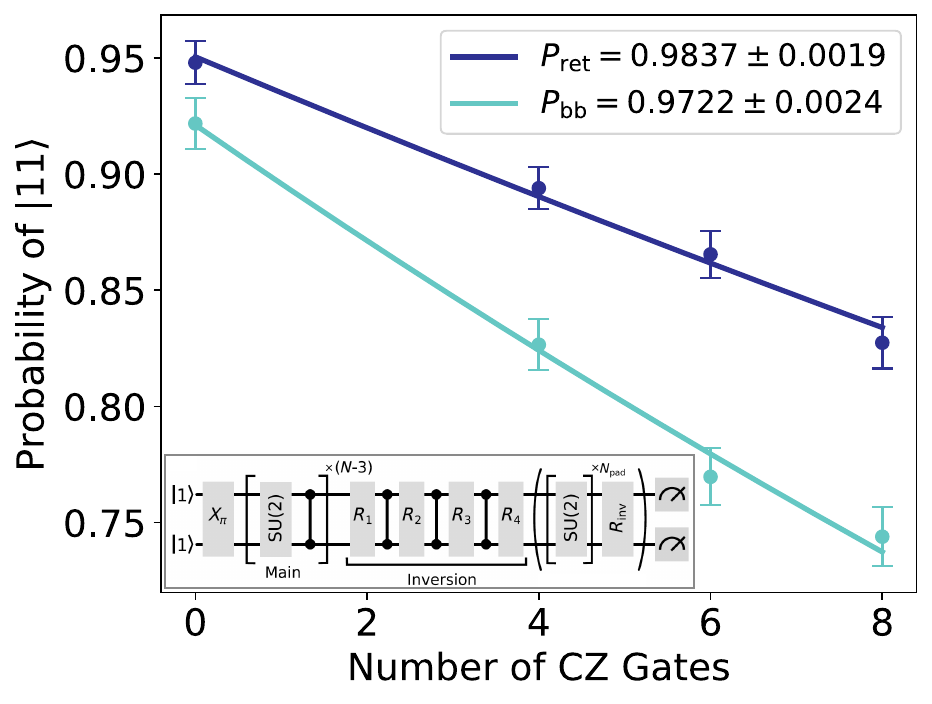}
    \caption{Randomized benchmarking of Rb-Cs $\sf CZ$ gates \rsubb{on a single pair of atoms}. Both fits are to the functional form $P_{\rm measure}=AP^n$, where $P_{\rm measure}$ is the measured probability of both atoms being present giving $P_{\rm ret}=P$ or both atoms being in the bright state after blowaway of the upper hyperfine level giving $P_{\rm bb}=P.$  The probability without any $\sf CZ$ gates due to state preparation and measurement errors is $A$, and $n$ is the number of $\sf CZ$ gates. \rsub{The RB circuit from \cite{Radnaev2025} is shown in the inset and the number of $\sf GR$ gates is padded to maintain a fixed number for all $n$.} 
    }
    \label{fig:RB_cz}
\end{figure}

{\it $\sf CZ$ Characterization --}
Entanglement between species is demonstrated with a $\sf CZ$ gate between individual Rb and Cs atoms, using a parameterized time optimal Rydberg gate protocol \cite{Jandura2022}, \rsub{wherein the Rydberg interaction Hamiltonian is presented}.  For $\sf CZ$ gates, Rb and Cs were excited to $\ket{63s_{1/2},m_j=-1/2}$ and $\ket{65s_{1/2},m_j=-1/2}$ Rydberg levels through 2 photon excitation.  These levels were chosen based on simulations showing little cross talk between these states and other excited states, which could complicate the blockade picture.  Ground-Rydberg Rabi rates of $\approx1.2$ MHz for both Rb and Cs are shown in Fig. \ref{fig:rydberg_rabi_data}. The figure also shows the blockade created with these two Rydberg levels.  In both \ref{fig:rydberg_rabi_data}c) and \ref{fig:rydberg_rabi_data}d), the Rb and Cs atoms in neighboring sites were initialized to their respective $\ket{1}$ states, a 415~ns Rydberg $\pi$ pulse was used to coherently excite the Cs atom, and a background microwave pulse ionizes the Cs Rydberg population, resulting in atom loss before measurement of the Cs atom. In \ref{fig:rydberg_rabi_data}d) the Cs excitation was preceded by a Rb excitation. In both cases, we sweep the laser frequency to find the Cs excitation resonance.  We find a shift in the Rydberg resonance of $2\pi\times12.01(22)$ MHz due to the blockaded excitation.
This shift was compared with Rydberg interaction calculations \cite{Miles2026SM} to verify the trap spacing of $5.85~\mu\rm m$.
The increase in excitation linewidth in \ref{fig:rydberg_rabi_data}d) can be attributed to shot-shot variations in the atom positions in each trap, which changes the distance between the atoms and the Rydberg interaction strength.  

\begin{figure*}[!t]
\includegraphics[width=0.95\textwidth]{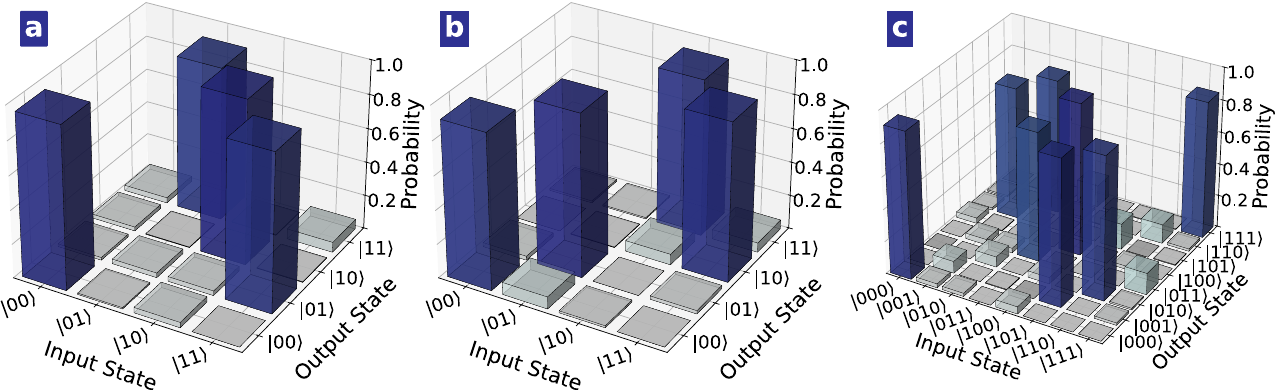}
    \caption{QND measurement results. \rsub{Two qubit states are $\ket{\rm Rb Cs}$; three qubit states are $\ket{\rm Rb Cs Cs}$} \textbf{a)} Two-atom QND with target qubit as Rb. \textbf{b)} Same as a), but with target qubit as Cs. \textbf{c)} Three-atom QND with target qubit as Rb. }
    \label{fig:qnd_results}
\end{figure*}

In order to apply a time-dependent phase to the Rydberg excitation to satisfy the gate protocol, fiber electro-optic modulators (EOMs) were installed on the 1005 and 1040 nm laser paths for Rb and Cs.  Calibration of the phase-voltage relationship for the EOM was done with a Ramsey experiment that used two $\frac{\pi}{2}$ pulses at 1005 nm (1040 nm) with a DC voltage applied on the EOM during the second pulse.  Scanning the voltage determined the level required to add $\phi=\pi$  to each laser pulse. The resulting calibration factors were used for implementing the gate.

A combination of global and individual qubit operations is required to show entanglement of Rb and Cs using Rydberg blockade. Similarly to previous work \cite{Graham2022,Radnaev2025}, individual addressing of the qubits uses focused Rydberg lasers by aligning the beams onto the appropriate sites with AOD scanners. 421 and 1005 nm lasers provide two photon excitation to Rydberg levels for Rb, while 459 and 1040 nm are used for Cs.

The fidelity of \rsubb{the} Rb-Cs $\sf CZ$ gate is characterized \rsubb{on a single pair} with randomized benchmarking (RB) \rsub{\cite{Radnaev2025}} with circuits composed of alternating $\sf CZ$ gates with global rotation ({\sf GR}) microwave gates of random phase and angle \rsub{on a single atom pair}.   Although the 6.8~GHz and 9.2~GHz microwave systems are independent, the same {\sf GR}  angles are applied to Rb and Cs, meaning the interleaved gates are chosen from $\text{SU}(2)$.  
The RB circuit requires proper single-qubit phase correction as part of each $\sf CZ$ gate.  In any same-species demonstration, the phase correction would have to be done by local $\sf R_z$ gates. In this work, for the purposes of a 
two-atom Rb-Cs demonstration, the phase correction on the $\sf CZ$ gates is done with {\sf GR} gates, since different {\sf GR} gates can be applied independently to each species of atom. Figure \ref{fig:RB_cz} shows RB data up to $\sf CZ$ gate depth of 8.  Since the maximum number of $\sf CZ$ gates in this RB is eight, a constant nine random {\sf GR} gates are applied in each gate depth, which follows the protocol in \cite{Radnaev2025}.  Five randomized circuits are used for the zero $\sf CZ$ gate depth case, while ten are used for the other cases, resulting in a total set of 35 circuits.  

\rsub{From the  data presented in Fig. \ref{fig:RB_cz} and the single qubit $\sf GR$ fidelity, we calculate a single-qubit state-preparation and measurement (SPAM) fidelity of $\mathcal{F}_{\rm SPAM}=0.984(7)$. This value is comparable to other reported SPAM fidelities in alkali atoms \cite{Radnaev2025}.} \rsubb{The separate contributions to the SPAM fidelity from state preparation or measurement were not independently determined.}

A hyperfine level-selective fluorescence measurement determines if both atoms remain in a bright state after the circuit, which we denote as $P_{\rm bb}$. This includes atoms in the correct final state $\ket{{\rm RbCs}}=\ket{00}$ as well as atoms that have leaked into lower hyperfine level states with $m_f\ne0$. 
Leakage errors occur with probability  $P_{\rm leak}$.  In our previous work \cite{Radnaev2025} we modeled the leakage probability to be $P_{\rm leak}=0.001$, where the detuning from the intermediate state was -2.1 GHz.  Since the work here is done with a smaller detuning of $\approx-1.4$ GHz, $P_{\rm leak}$ is conservatively estimated to be 0.002.  In order to properly account for losses, the same set of circuits is run with an occupancy measurement, indicating if an atom was lost during the circuit, most likely due to incomplete transfer  of Rydberg atoms back to the ground state.   Each set of data was fit with a decaying exponential with an asymptote fixed to zero, thus allowing us to extract the per-gate occupancy or retention probability $P_{\rm ret}=1-P_{\rm loss}$ and $P_{\rm bb}$. The probability of observing the bright state given that the atoms were not lost is   $P_{\rm bb|ret}=P_{\rm bb}/P_{\rm ret}$.  
The gate fidelity is then found from \cite{Radnaev2025}
\begin{equation}
   \mathcal{F} = P_{\rm ret}(1-P_{\rm leak})\left(1-\frac{3}{4}\sigma \right)
\label{eq:occupancy_model}
\end{equation}
with the probability of a depolarizing error given by $\sigma=(1-P_{\rm leak}-P_{\rm bb|ret})/(1-P_{\rm leak})$.
The data in Fig. \ref{fig:RB_cz} with Eq. (\ref{eq:occupancy_model})  
results in a $\sf CZ$ fidelity of  ${\mathcal F}=0.975(2)$. This is considerably higher than the previously measured Bell fidelity for Rb and Cs of 0.69(3) \cite{Anand2024}, or the highest reported entangling gate fidelity between heteronuclear atom pairs of 0.73(1) \cite{YZeng2017}. \rsubb{Similar performance was seen on other atom pairs, but quantitative fidelity benchmarking was not performed.}

\rsub{Our measured gate fidelity matches well with a detailed  error budget \cite{Miles2026SM}. The leading error channels are scattering from the intermediate state in the two-photon Rydberg excitation, inhomogeneous laser pulse power, and decay from the Rydberg state during the gate. In \cite{Miles2026SM}, we provide pathways towards mitigating these and other error sources, in order to reach  $\mathcal{F}>0.997$ with realistic experimental improvements.}

{\it 
Quantum Non-Demolition Measurements --}
Two- and three-atom quantum non-demolition (QND) circuit measurements were performed to further show entanglement of Rb-Cs and to demonstrate the use of two-species for mid-circuit syndrome measurements. A Rb-Cs gate was also previously used for a two-atom QND measurement in \cite{Anand2024}.   The circuits and atom geometry used for QND demonstrations are shown in Fig. \ref{fig:setup}. Figure \ref{fig:qnd_results} shows state measurement probabilities for the two- and three-atom QND measurements. The three-atom experiment corresponds to a {\sf ZZ} stabilizer measurement, which is a weight-2 parity check, as appears on the boundary of the rotated surface code. The same circuit also appears in pre-compiled versions of Shor's algorithm \cite{Rines2025}.   Two-atom measurements were run with both Rb or Cs as the ancilla qubit; three-atom measurements used Rb as the ancilla.

Success probabilities and associated uncertainties of the 2- and 3-atom QND measurements are obtained from the marginal Dirichlet posterior distributions of the events corresponding to correct outcomes, assuming a uniform prior. The raw count data are given in \cite{Miles2026SM}. Averaging over all initial states in the  two-qubit experiments, we determine syndrome measurement fidelity of $\mathcal{F_{\mathrm{QND}}}=0.937(8)$ when using Rb as the target qubit and $\mathcal{F_{\mathrm{QND}}}=0.929(9)$ when using Cs as the target. For three-qubit syndrome extraction, we calculate a syndrome measurement fidelity of $\mathcal{F_{\mathrm{QND}}}=0.865(17)$.
Since the three-qubit circuit involved multiple atoms of the same species, input state preparation used a combination of $\sf R_z$ and {\sf GR} gates.  The circuit demonstrates that if the parity of Cs states is odd (one in $\ket{1}$ and one in $\ket{0}$) the Rb atom will experience a bit-flip. 

In these demonstrations, all atoms are measured after the circuit for characterization. However, to execute error correction, only the ancilla qubits would be measured to determine if an error occurred on the data qubit. This  requires independent and crosstalk-free Rb and Cs readouts, which we demonstrate in \cite{Miles2026SM}.

{\it Outlook --} We have demonstrated \rsubb{and benchmarked} an interspecies Rydberg gate \rsubb{on a selected pair of}  Rb and Cs atoms that achieves a fidelity of $\mathcal F=0.975.$ Simulations predict that the interspecies gate fidelity can be improved to better than 0.997 with realistic improvements to the experimental apparatus. The ability to measure ancilla qubits without disturbing neighboring data provides a path to measurement based error correction without incurring additional overhead from atom transport to a measurement zone, shelving, or hiding operations. The inter-species gate is also of interest for implementing new variants of the surface code  \cite{Eickbusch2025}.

{\it Acknowledgement --} 
We thank M. Bergdolt for assistance in developing the experimental apparatus. SAN and MS received support from the US National Science
Foundation under Award 2016136 for the QLCI center
Hybrid Quantum Architectures and Networks \rsubb{and by ARO
contract W911NF2410382}.

\bibliography{qc_refs,optics,saffman_refs,rydberg,atomic,thispaper}

\pagebreak

.

\newpage

\setcounter{page}{1}
\setcounter{section}{0}
\setcounter{figure}{0}
\setcounter{equation}{0}
\setcounter{table}{0}

\renewcommand{\thepage}{SM.\arabic{page}}
\renewcommand{\theequation}{SM.\arabic{equation}}
\renewcommand{\thesection}{SM.\arabic{section}}
\renewcommand{\thefigure}{SM.\arabic{figure}}
\renewcommand{\thetable}{SM.\arabic{table}}

\onecolumngrid

\noindent
{ \bf Supplemental Material for \\
\Large \center Qubit syndrome measurements with a high fidelity Rb-Cs Rydberg gate}


\section{Qubit characterization and coherence}

\begin{figure}[!b]
    \centering
    \includegraphics[width=0.5\textwidth]{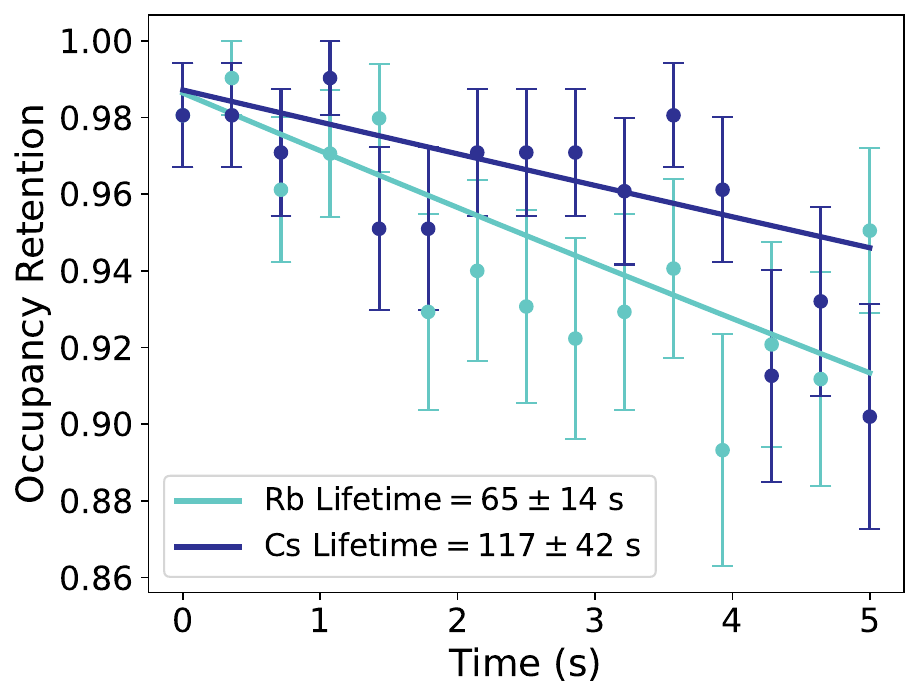}
    \caption{Lifetime of single atoms while in the traps.  Atoms are prepared in $\ket{1}$ and then left in the trap for up to 5 seconds before an occupancy readout.}
    \label{fig:atom_lifetime}
\end{figure}

The qubit array is established in a glass cell fabricated at Infleqtion. Figure \ref{fig:atom_lifetime} shows trapping lifetimes of single Rb and Cs atoms. The lifetime is limited by background collisions with untrapped atoms or background gas, and possibly heating due to trap laser noise. The Cs lifetime is almost twice longer than that of Rb which is consistent with the deeper trap depth for Cs, but we have not attempted a quantitative comparison with theoretical lifetime estimates which depend on the background pressure and molecular composition.  The lifetimes are much longer than the qubit coherence times and are not a limiting factor at the fidelity levels presented in this paper.

\begin{figure}[!t]
    \includegraphics[width=\textwidth]{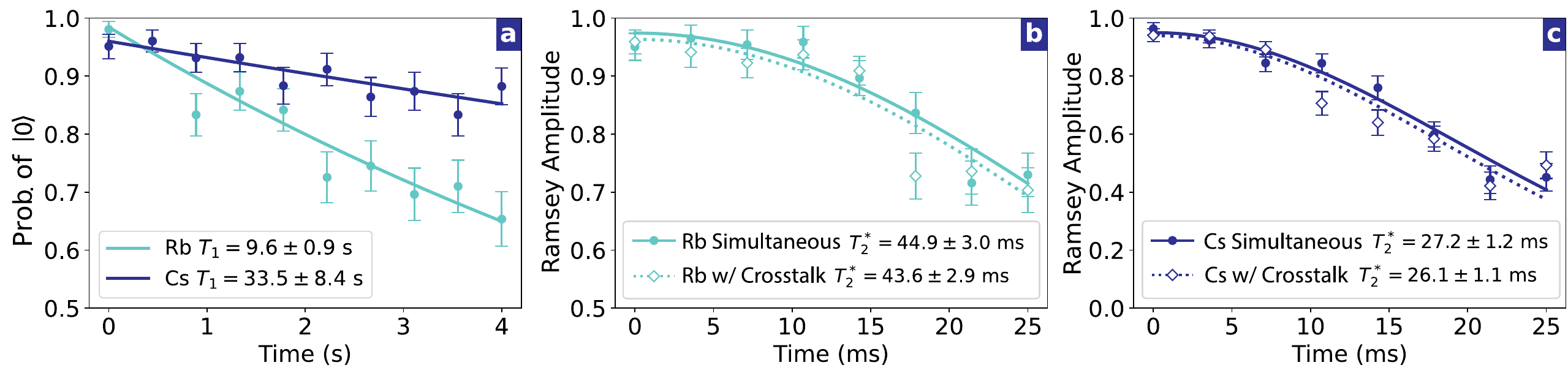}
    \caption{
        \rsub{
            Characterizations of qubit coherence. Each experiment has one Rb and one Cs qubit present.
            \textbf{a)} Population leakage from $\ket{0}$ to $\ket{1}$, measured simultaneously, yielding $T_{1,\rm Rb}=9.6(9)$ s and $T_{1,\rm Cs}=33.5(8.4)$ s.
            \textbf{b)} Rb microwave Ramsey. Measured simultaneously with Cs yields $T_{2, \rm Rb}^*=44.9(3.0)$ ms.  Measured with crosstalk from Cs readout light on during the free evolution yields $T_{2, \rm Rb}^*=43.6(2.9)$ ms.
            \textbf{c)} Cs microwave Ramsey. Measured simultaneously with Rb yields $T_{2, \rm Cs}^*=27.2(1.2)$ ms.  Measured with crosstalk from Rb readout light on during the free evolution yields $T_{2, \rm Rb}^*=26.1(1.1)$ ms.
        }
    }
    \label{fig:t1t2}
\end{figure}

\rsub{Coherence measurements shown in Fig \ref{fig:t1t2} were performed on neighboring single Rb and Cs atoms in the 49 site array, with both atoms present to demonstrate the isolation between the Rb and Cs registers.}

\rsub{$T_1$ values were measured by simultaneously preparing and observing population leakage on both species.  The data shown is for the experimentally worse leakage direction, with atoms prepared in the lower state $\ket{0}$ and the population leaking to the upper hyperfine manifold was measured as a function of wait time.  The significantly shorter $T_1$ time for Rb than for Cs was traced to a small leakage of Rb D2 repump light from an acousto-optic switch.  Since the $T_1$ time for both species is six orders of magnitude longer than the gate and measurement times reported here, the $T_1$ coherence is also not a limiting factor in the reported results.}
 
\rsub{For $T_2^*$ we perform Ramsey measurements simultaneously on the single neighboring Rb and Cs atoms. The ``simultaneous" measurements shown in Fig \ref{fig:t1t2} (b) and (c), are from the same experiment.  We also performed two Ramsey experiments that had readout light for one species on during the free evolution gap.  This spoils the coherence on that species, but gives us little to no ``crosstalk" on the measured species.  The ``simultaneous" and ``crosstalk" measurements differ by less than one standard deviation (standard scores $z_{\rm Rb}=0.32$, $z_{\rm Cs}=0.71$) indicating no statistically significant difference between the samples.  We attribute this to the spectral separation of the resonant transitions between the two atomic species of 25 THz between the closest D lines.  These data indicate decoupled data and ancilla registers, a necessary characteristic for crosstalk-free mid-circuit measurements for error correction.  The observed $T_2^*$ values are primarily dependent on atom temperature which leads to differential light shifts of the qubit states as the atom moves in the optical trap, and magnetic field noise.  The measured values are consistent with magnetic noise of approximately 5 mG together with the temperatures measured by trap drop and recapture \cite{Saffman2011}.}

Trap depths for Rb and Cs were calculated by analyzing the light shift on the $\ket{0}$ and $\ket{1}$ states measured with microwave spectroscopy.  The microwave resonance was analyzed at different trap intensities and based on the shift in resonance the exact trap intensity was calculated.  Radial and axial trap frequency measurements were also taken at different trap powers, and a combination of the trap depth and trap frequencies led to a calculated beam waist.  

\begin{table}[!t]
\centering
\caption{Rydberg laser beam waists ($1/e^2$ intensity radius) in the atom plane ($\mu$m). These measurements are from intentionally misaligning the lasers from the targeted atom, and performing a Ramsey-Stark measurement on the atom (Ramsey sequence with laser pulse inserted between the $\pi/2$ rotations to differentially Stark shift the qubit states). Multiple frequencies were used per laser, and this data was fit against the frequency of the AOD that was scanned. The frequency could be scaled to atomic position using the nominal 5.85 $\mu$m spacing between adjacent sites which was determined from comparison of measured Rydberg interaction strength (Fig. 3 in the main text)  with simulations in Fig. \ref{fig:63-65-pair-states}.  }
\label{tab:beam_sizes}
\Large
\begin{tabular}{|c|c|c|c|c|}
\hline
 & 421 & 459 & 1005 & 1040 \\
\hline
    Horizontal  & 3.48  & 5.10  & 7.44  & 3.76  \\
    Vertical    & 3.42  & 3.85  & 7.11  & 3.66 \\
    Average     & 3.45  & 4.48  & 7.28  & 3.71 \\
    \hline
\end{tabular}
\end{table}
\normalsize

\section{Rydberg lasers}

\subsection{Laser System}

The two sets of lasers for Rydberg state excitation are implemented using Vexlum VALO (421 nm (internally doubled 842 nm), 1005 nm, 1040 nm) and MSquared Solstis \& doubler  (918 \& 459 nm) systems.  The NIR lasers are locked to a single, 4-bore Ultra-Low-Expansion glass (ULE) (Stable Laser Systems), high finesse cavity, with each bore having a different set of mirrors for the specific wavelength (842 nm, 918 nm, 1005 nm, 1040 nm). A PDH scheme is used to lock these lasers to the cavities allowing for line narrowing and frequency stabilization, with offset AOMs used to adjust the frequency of the laser from the nearest cavity resonance to the desired frequency for the Rydberg resonance. \rsub{Beam sizes at the atoms are presented in Table \ref{tab:beam_sizes}}.

\subsection{Noise Measurements and Characterization}

The spectral noise of the Rydberg lasers has a detrimental effect on the fidelity of the Rydberg Rabi oscillations. We utilize the self-heterodyne approach presented in \cite{XJiang2023}  for measuring and characterizing the spectral noise. As in that work we use a model where the laser noise can be approximated by a white noise floor with servo bumps at specific frequencies so  the frequency noise power spectral densities (PSD) of the laser systems are  linear combinations of a white noise contribution with servo-bumps added from the locking mechanisms. The frequency PSD ($S_{d\nu}$) for white noise is modeled as $S_{d\nu,0} = h_0$, with $h_0$ the amplitude of the white noise and the zero subscript indicating the white noise component. In terms of phase noise ($S_{\phi}(f) = S_{d\nu}/{f^2}$), this is simply $S_{\phi, 0}(f) = \frac{h_0}{f^2}$.

The servo-bump frequency noise PSDs are modeled as pairs of symmetric gaussian peaks, 

\begin{equation}
    \label{eq.sdnu-bump}
    S_{d\nu,j}(f) =h_j\left[\exp\left({-\frac{(f-f_j)^2}{2\sigma_j^2}}\right)+\exp\left({-\frac{(f+f_j)^2}{2\sigma_j^2}}\right)\right]
\end{equation}
where $h_j$ is the amplitude of the $j^{th}$ servo bump, $f_j$ is its center frequency, and $\sigma_j$ is the gaussian width. In the phase noise space,
\begin{equation}
    \label{eq.sphi-bump}
    S_{\phi, j}(f) =\frac{h_j}{f_j^2}\left[\exp\left({-\frac{(f-f_j)^2}{2\sigma_j^2}}\right)+\exp\left({-\frac{(f+f_j)^2}{2\sigma_j^2}}\right)\right]
\end{equation}

With this, the spectrum of the laser phase noise is modeled as a linear combination of these contributions,
\begin{equation}
    \label{eq.servo-summing}
    S_{\phi, \rm total} = S_{\phi,0}(f)+\sum_j S_{\phi,j}(f).
\end{equation}

We use a self-heterodyne technique to extrapolate the spectral noise of the laser systems. As discussed in reference \cite{XJiang2023}, the modeled white noise and servo bumps subjected to this measurement scheme would theoretically be measured as the analytic functions $S_{i,0}(f)$ and $S_{i,j}(f)$, respectively:
\begin{equation}
    \label{eq.si-white}
    S_{i,0}(f)= \frac{2h_0}{f^2+(2\pi h_0)^2}+e^{-4\pi^2h_0t_d}\Biggl(\delta(f) 
     -\frac{2h_0}{f^2+(2\pi h_0)^2}\left[\cos{(2\pi f t_d)}+\frac{2\pi h_0}{f}\sin\left(2\pi ft_d\right)\right]\Biggr)
\end{equation}
\begin{equation}
    \label{eq.si-bump}
    S_{i,j}(f)\approx\frac{4h_j}{f_j^2}\sin^2\left(\pi f t_d\right)\Biggl[\exp\left({-\frac{(f-f_j)^2}{2\sigma_j^2}}\right) +\exp\left({-\frac{(f+f_j)^2}{2\sigma_j^2}}\right)\Biggr]
\end{equation}
and combining, the normalized ($\int_{-\infty}^{\infty}S_{i}(f)df=1$) self-heterodyne PSD data is modeled as
\begin{equation}
    \label{eq.servo-fitting}
    S_{i}(f) = S_{\rm dark} + S_{i, 0}(f)+\sum_j S_{i,j}(f)
\end{equation}
where $S_{\rm dark}$ is a fit parameter to model the dark noise measured by the spectrum analyzer.

\begin{figure}[!t]
 \includegraphics[width=0.85\columnwidth]{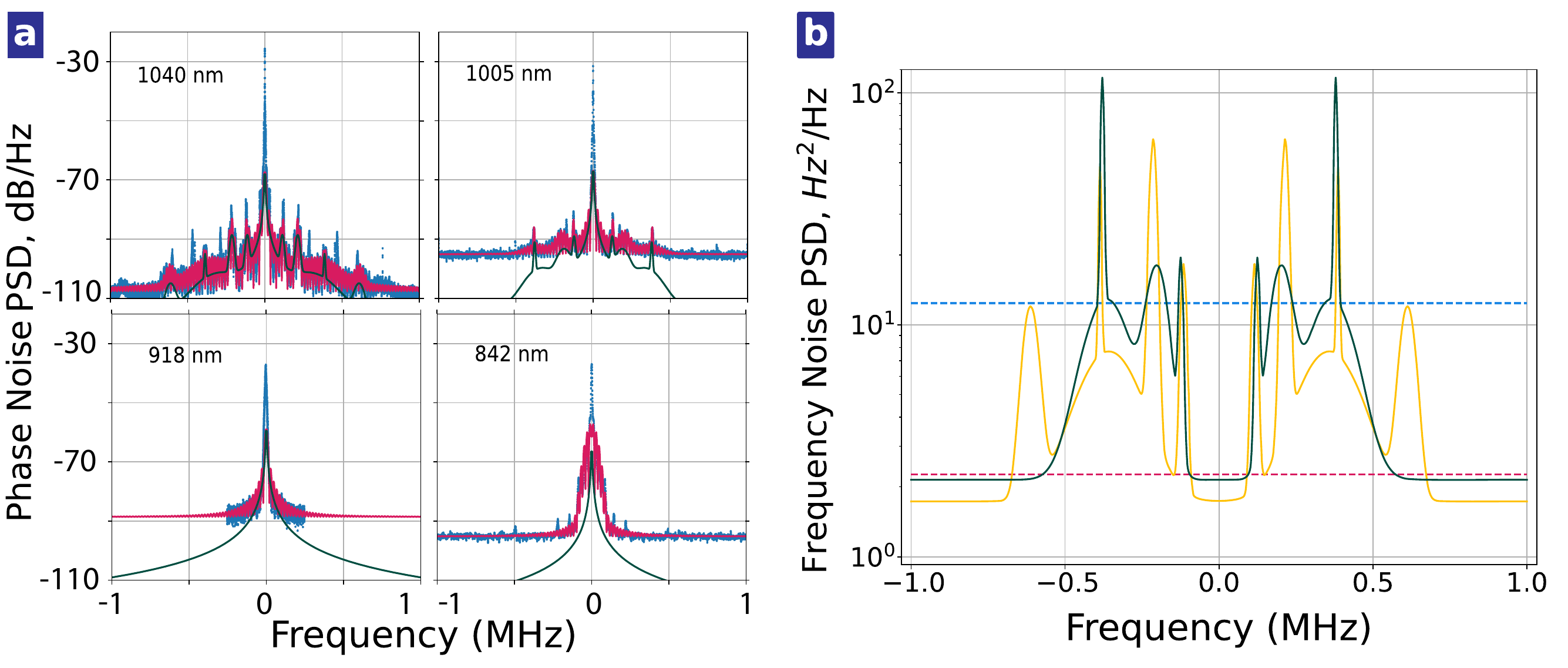}
    \caption{Modeling the Rydberg laser noise spectral densities. \textbf{a)} Measured Rydberg laser phase noise using a self-heterodyne system with a 10 km fiber delay for each Rydberg laser. Normalized data (blue points) are fit to $S_{i}(f)$ (red lines) resulting in a modeled fit phase noise with dark noise extracted $S_\phi(f)$ (green lines). Not all servo bumps are fit since smaller width bumps contribute much less but increase the base white-noise level. Additionally, since the dark noise of the 842 nm and 918 nm lasers is higher due to SNR, they are only fit to white noise distributions. \textbf{b)} Modeled frequency noise spectral densities for 842 nm (red, dashed), 918 nm (blue, dashed), 1005 nm (yellow, solid), and 1040 nm (green, solid). The solid lines show the modeled fit with servo bumps while the dashed lines indicate that only the white noise is being presented.}
    \label{fig:spectral_noise_fit_and_data}
\end{figure}

Figure \ref{fig:spectral_noise_fit_and_data} presents this analysis for each of the Rydberg lasers, presenting both the fitting and the total modeled frequency noise PSD. For the SHG doubled lasers (421/842 nm and 459/918 nm), an insufficient amount of laser power is sent to the self-heterodyne system limiting the ability to see the effects of the laser noise below the dark noise level. As such, only the 1005 nm and 1040 nm lasers have enough signal-to-noise ratio to accurately measure the servo-bump noise level. Of note, the white noise level ($h_0$) of the VECSEL based (Vexlum) Rydberg lasers (842 nm, 1005 nm, 1040 nm) is $\approx2$ Hz$^2$/Hz while the Ti:Sa based (MSquared) (918 nm) is $\approx12$ Hz$^2$/Hz.  

The spectral noise of the Rydberg lasers is a key contributor to the error accrued from driving a Rydberg Rabi oscillation. Reference \cite{XJiang2023} demonstrates that the error for a full ground-Rydberg Rabi oscillation through an intermediate state can be well approximated by adding the error of the independent single-photon Rabi rotations at the Rabi rate of the full ground-Rydberg rotation. We use this framework to calculate the expected $2\pi$ Rabi rotation error of the Rydberg gates caused by the spectral noise of the Rydberg lasers.

The Rabi rotation error, $\mathcal{E}$ accumulated from an $N\pi$ Rabi rotation with Rabi frequency $\Omega_0$ from a laser with frequency noise PSD $S_{d\nu}(f)$ is
\begin{equation}
    \mathcal{E}=4\pi^2\int_0^\infty df S_{d\nu}(f)\frac{\Omega_0^2\left[1-(-1)^{2N}\cos{\left(4\pi^2Nf/\Omega_0\right)}\right]}{\left(\Omega_0^2-4\pi^2f^2\right)^2}
    \label{eq.Rabi_error_vs_f_noise}
\end{equation}

\begin{figure}
    \centering
    \includegraphics[width=0.5\columnwidth]{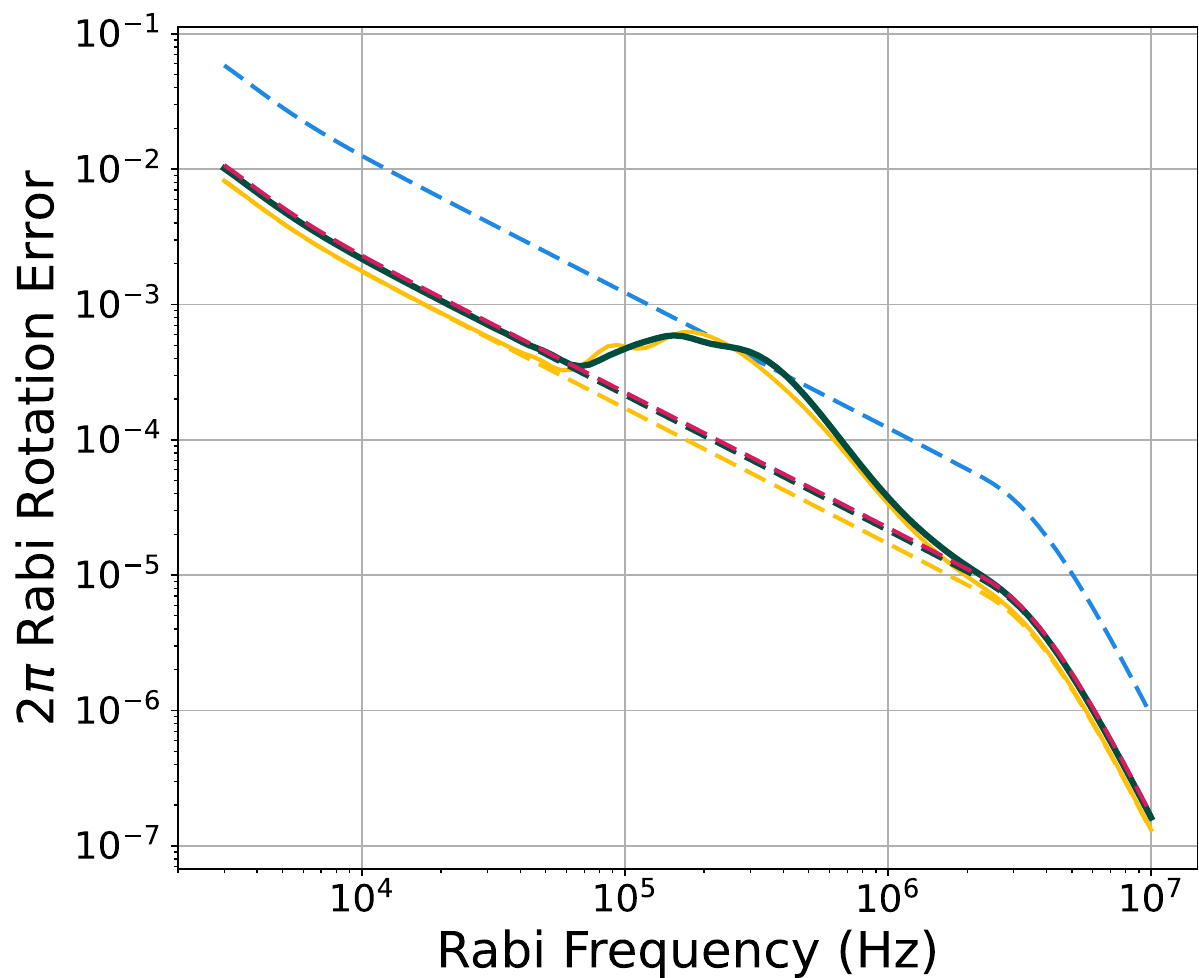}
    \caption{Predicted $2\pi$ Rabi rotation error as a function of Rabi frequency using Eq.  (\ref{eq.Rabi_error_vs_f_noise}) for each Rydberg laser: 842 nm (red), 918 nm (blue), 1005 nm (yellow), and 1040 nm (green). Dashed lines present error with a white-noise only model while solid lines (1005 nm and 1040 nm only) present the error contribution from the (measurable) servo-bumps. These errors can be added linearly to extrapolate the total error associated with the laser noise spectrum for the two-photon ground-Rydberg $2\pi$ Rabi rotation. }
    \label{fig:Rydberg_gate_error}
\end{figure}

This integration can be performed on the measured noise from the self-heterodyne spectrum of each Rydberg laser to extrapolate their independent contributions to the Rabi rotation errors. Figure \ref{fig:Rydberg_gate_error} presents the expected error accumulation from a single-photon $2\pi$ Rabi rotation as a function of the Rabi frequency for each of the Rydberg lasers. For a Rabi frequency of $\Omega_0 =2\pi\times  1$ MHz, each Rydberg lasers contributes to an error less than $10^{-3}$ and for a two-photon ground-Rydberg $2\pi$ Rabi rotation, adding these errors linearly also leads to an error less than $10^{-3}$.  This error is currently negligible compared to the errors accounted for in Sec. \ref{sec.error_budget} and has therefore not been included in the full error analysis. For future, higher fidelity gates, reduction of laser noise will be important for achieving the highest possible fidelity. 

\subsection{Auto-Relock system}

To maintain long-term stability of the Rydberg lasers, an auto-relock system is implemented to disable, re-tune and activate the system as the changing lab conditions can cause laser mode and lock instabilities. This scheme involves acquiring input from a Fizeau wavemeter for coarse frequency tuning, CCD cameras to measure cavity transmission beam profiles for mode-analysis, and photodiodes monitoring the ULE cavity transmission intensity for fast analysis of the laser lock status. Additionally, high bandwidth ($>1\rm\  GHz$) photodiodes are used to measure the spectral mode composition of the Vexlum lasers, which have a tendency to become multi-mode when subject to environmental fluctuations. 

The general scheme involves the system deactivating the feedback loop integrators, tuning the laser back to an optimal operating condition, then reactivating the integrators to resume the high-fidelity locking state of the system. Depending on the cause of the unlock event, determined through the myriad of the aforementioned metrics, the system tunes multiple features of the laser. Scanning the offset of the cavity's piezo actuated mirrors allows for basic adjustments while optimizing the temperature of the cavity's etalon allows for correction of multi-mode behaviors often found in the Vexlum laser systems. 

These auto-relocking schemes bring laser up-time to $>99.9\%$, with quick relock and tuning operations performed in just a few seconds (occurring up to a few times a day) while the more robust temperature optimizations are performed in minutes (occurring on a weekly time-scale).

\section{Rydberg interactions}

\begin{figure}[!t]
    \centering
    \includegraphics[width=.85\columnwidth]{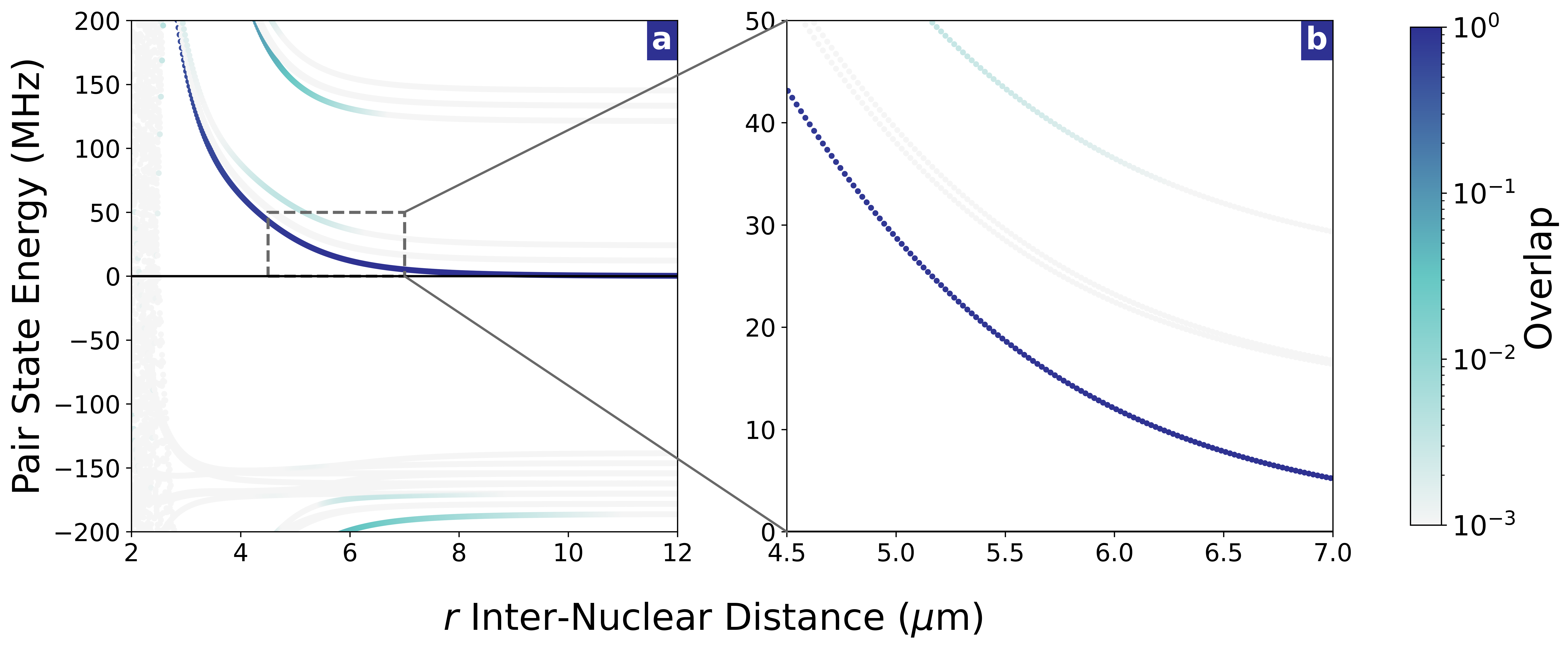}
    \caption{\textbf{a)} Rydberg atom pair potential for $n_{Rb}=63$, $n_{Cs}=65$. \textbf{b)} Zoom of panel a). }
    \label{fig:63-65-pair-states}
\end{figure}

\begin{figure}[!t]
    \centering
    \includegraphics[width=0.85\columnwidth]{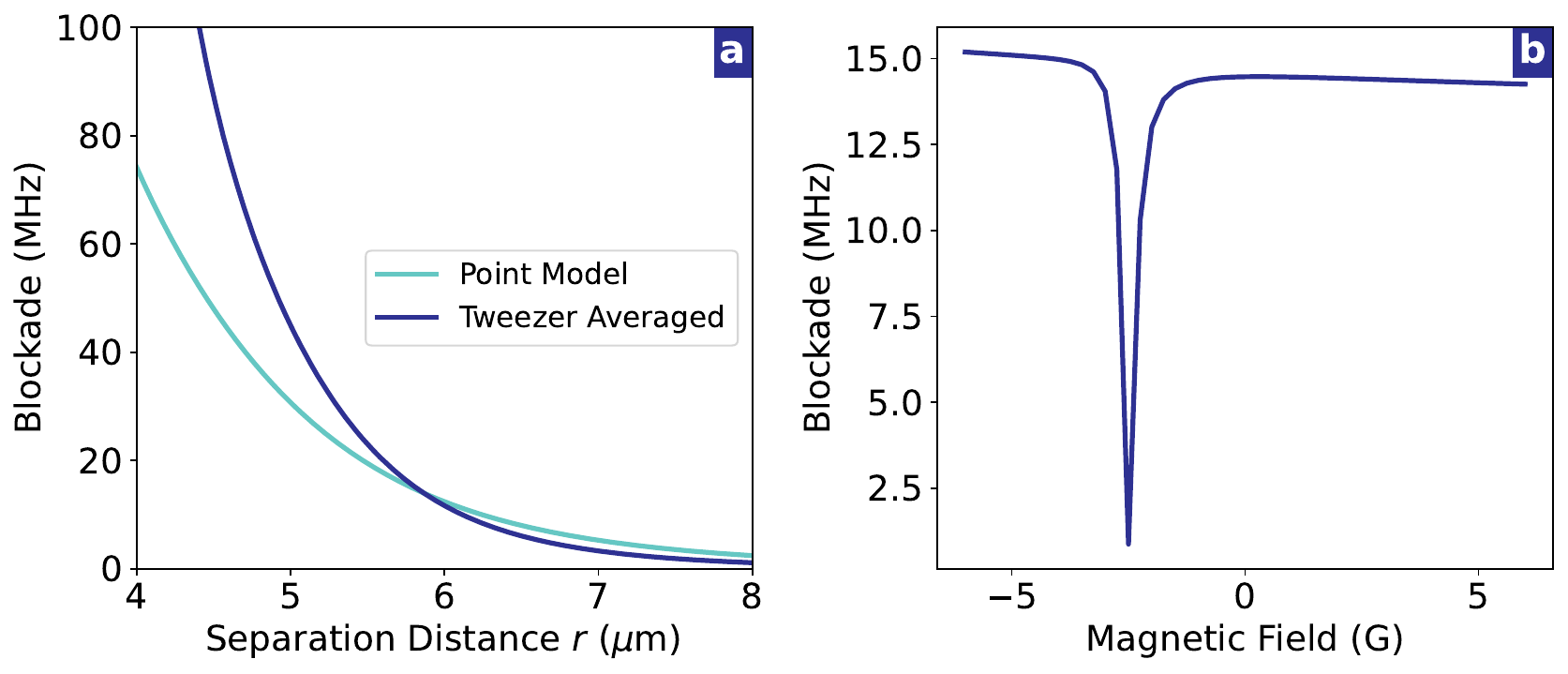}
    \caption{\textbf{a)} Rydberg blockade, modeled both with point model and averaged over positional variation in tweezers. For the latter, ``Separation Distance" refers to the center-to-center distance of the individual laser tweezers. \textbf{b)} Variation of blockade with applied magnetic field, calculated for point model $r=5.85 \ \mu\rm m$.}
    \label{fig:blockade-in-tweezer}
\end{figure}

Rydberg pair state interaction energies were calculated with the Alkali Rydberg Calculator (ARC) 
\cite{Sibalic2017}.
 Pair states were calculated for the target state $\ket{n_{\rm Rb} s_{1/2}, m_j=-\frac{1}{2}}\otimes \ket{n_{\rm Cs} s_{1/2}, m_j=-\frac{1}{2}}$. The basis for the calculation is described by $\Delta n=\Delta \ell=4,$  keeping only states within $\Delta E_{\rm max}=25$ GHz from the infinite-distance pair state energy. Atom geometry, as specified by the experimental setup is set as $\theta=\frac{\pi}{2},\phi=0$. A magnetic field of 4.25 G was used in the calculation, matching experimental values. The pair state for the interaction used in this letter, $(n_{\rm Rb},n_{\rm Cs})=(63,65)$, is shown in Figure \ref{fig:63-65-pair-states}.
Pair potential eigen-channels can be collapsed to a single Rydberg-Rydberg blockade channel via a direct summation \cite{Walker2008}. The results for this are shown in Fig. \ref{fig:blockade-in-tweezer} along with the dependence of the blockade on the bias magnetic field. For the field of 4.25 G used here there is only a very weak dependence on the field strength. 

Since the distribution of the blockade over atom separation distances is non-uniform, the mean blockade for a given tweezer separation is, in general, not equal to that of two atoms localized at the center of the same tweezers.
With the RMS atom positional spreads (post-adiabatic cooling) presented in the main text, we generate distributions of Rydberg blockade, limited by finite atom localization uncertainty. Mathematically, the average blockade of atoms initialized in tweezers is 
\begin{equation}
    \braket{B}=\int\int
    d^3r_1d^3r_2\,
    \rho_1(\vec{r_1})\rho_2(\vec{r_2})B\left(|\vec{r_2}-\vec{r_1}|\right),
\end{equation}
where $\rho_1$ and $\rho_2$ are the normalized spatial distributions of the two individual atom positions. The tweezer averaged blockades for the candidate states are shown in Fig. \ref{fig:blockade-in-tweezer}. We note that these blockade figures differ from the center-of-tweezer model, as stochastic processes initializing atoms closer than their mean separation distances adds blockade terms of large magnitude to the average. Additionally in this figure, we show the dependence of the blockade on the magnitude of the applied magnetic field. The sharp peak corresponds to a region where the pair state potential crosses from positive to negative energy at the separation distance of interest. As such, we operate in a region far from this resonance to avoid  blockade suppression.

We experimentally measure the blockade by exciting rubidium to the Rydberg state and attempting a cesium Rydberg excitation, sweeping the frequency of the cesium excitation pulse. We measure a blockade of $12.01\pm0.22$ MHz.
This implies a tweezer separation of $5.85\pm0.02$ $\mu\rm m$.

\section{Gate Pulse}
\begin{figure}
    \centering
    \includegraphics[width=0.5\linewidth]{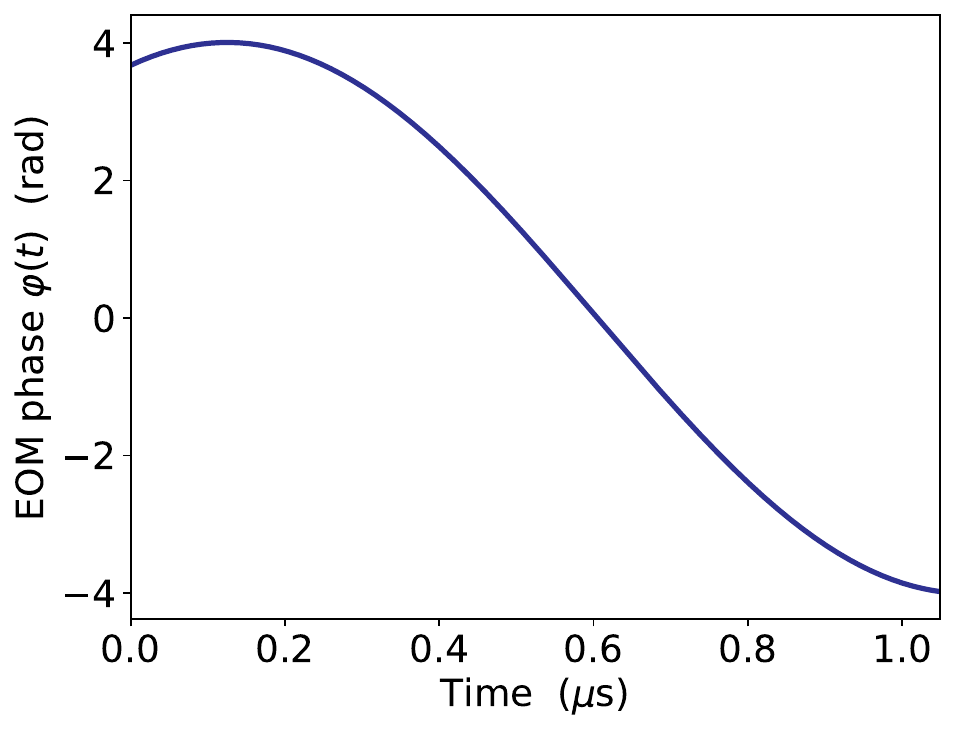}
    \caption{\rsub{Parameterized time-optimal pulse used for the $\sf CZ$ gate.}}
    \label{fig:TO_phase_profile}
\end{figure}

\rsub{We use a parameterized Time-Optimal-type gate with a constant intensity amplitude and a time-varying phase. The phase has a sinusoidal functional form.
We apply the gate pulses to the atoms and use gradient descent methods to tune the gate duration, detuning, and phase offset using gate infidelity as our loss function for feedback. This pulse is shown in Fig. \ref{fig:TO_phase_profile}.
The parameters used for the gate are as follows:
\begin{itemize}
    \item Total gate time: 1.036 $\mu$s
    \item Average Rabi frequency: $2\pi\times 1.215$ MHz.
    \item Detuning: $+2\pi\times 1.565$ MHz
    \item Modulation frequency: $2\pi\times 0.528$ MHz
    \item Modulation depth: 4.00 rad
\end{itemize}
}

\rsub{Here we describe the methodology used to calculate the single-qubit state preparation and measurement (SPAM) fidelity, using the data taken in the $\sf CZ$ randomized benchmarking experiments. Since each experiment has the same number of $\sf GR$ gates (9), we can decompose the total fidelity for each point, as a function of the number of $\sf CZ$ gates applied $n$: 
\begin{equation}
    \mathcal{F}_i(n)=\mathcal{F_{\rm SPAM}}\mathcal{F}_{\sf{ GR}, \rm{Rb}}^9\mathcal{F}_{\sf{GR}, \rm{Cs}}^9\mathcal{F}_{\sf CZ}^n,
\end{equation}
for $i\in\{\rm ret, bb\}$ (retention or bright-bright). With $\F_i=A_ip_i^n$, the two-qubit SPAM fidelities are given
\begin{equation}
    \F_{\rm{SPAM},i}=\frac{A_i}{\mathcal{F}_{\sf{ GR}, \rm{Rb}}^9\mathcal{F}_{\sf{GR}, \rm{Cs}}^9},
\end{equation}
From this formula, we derive $\F_{\rm SPAM, ret}=0.957(9)$ and $\F_{\rm SPAM, bb}=0.927(11)$. We consider the SPAM fidelity of 
$\F_{\rm SPAM,2q}= \F_{\rm SPAM, bb}/\F_{\rm SPAM, ret}$, as the two-qubit state preparation and measurement fidelity, conditioned on atom survival. We determine this to be $\F_{\rm SPAM,2q}=0.969(14)$. Finally, with an assumption that the SPAM errors are independent on Rb and Cs registers, the single-qubit SPAM fidelity, conditioned on atom survival is $\F_{\rm SPAM}=0.984(7)$.} \rsubb{The separate contributions to the SPAM fidelity from state preparation or measurement were not independently determined.}

\rsub{For completeness we have accounted for  the fidelity of the global rotation microwave gates  ${\mathcal F}_{{\sf GR} ,\rm{Rb}},  {\mathcal F}_{{\sf GR}, \rm{Cs}}$ even though some of the benchmarking measurements were taken using virtual phase tracking instead of GR gates. This has very minor influence on the SPAM  results since the GR fidelities were $> 0.9996$ (see Fig. 2 in the main text) and $(0.9996^9)^2 =0.993$. }

\section{Adiabatic Cooling}

\rsub{
Adiabatic reduction of the trap depth is expected to lower the atom temperature according to $T_{\rm f}=T_{\rm i} \sqrt{U_{\rm f}/U_{\rm i}}$ \cite{Tuchendler2008}.  Given our initial atom temperature measurements, this scaling predicts final atom temperatures of $T_{\rm f}= 4.0,\ 4.5 ~\mu\rm K$ for Rb, Cs, which is within about 20\% of the observed final temperatures. }

\rsub{The root mean square spreads of the atom position in the radial and axial directions are 
$\sigma_\rho=(w/2)\sqrt{k_{\rm B}T/U}$, $\sigma_z=(\pi w^2/\sqrt2 \lambda)\sqrt{k_{\rm B}T/U}$ with $\lambda$ the wavelength of the trap light \cite{Saffman2005a}. Before the adiabatic ramp we calculate initial spreads of 
$\sigma_{\rho}, \sigma_z=0.10,\ 0.71~\mu\rm m$ and $0.079,\ 0.56~\mu\rm m $ for Rb and Cs respectively. The calculated  localization parameters at the final conditions are $\sigma_{\rho}, \sigma_z=0.18,\ 1.3~\mu\rm m$ and $0.15,\ 1.1~\mu\rm m $ for Rb and Cs. We obtain better final localization than predicted by the adiabatic scaling of 
$\sigma_{\rm f}/\sigma_{\rm i}=(U_{\rm i}/U_{\rm f})^{1/4}$ 
since we observe slightly better than $\sqrt{U_{\rm f}/U_{\rm i}}$ temperature reduction.
}

\section{Error Budget and Simulations}
\label{sec.error_budget}

We evaluate the {\sf CZ} error budget using a Monte Carlo simulation which is generalized to dual species from our previous work \cite{Radnaev2025}.  This simulation follows the evolution of a three-level ($\ket{0}, \ket{1}, \ket{r}$) state vector under a time-dependent non-Hermitian Hamiltonian.  The non-Hermitian terms are included to model scattering and decay out of the three-level system.  The simulation models a minimal test circuit that would produce a Bell state for a good gate, and the error is calculated as $1-\left|\bra{\psi_{\rm test}} \psi_{\rm Bell}\rangle\right|^2$.  The simulation benefits from having direct access to the final quantum state, rather than the stochastic measurements made on our real system, and can therefore assess the fidelity on every shot.  The Hamiltonian changes in time due to the applied laser fields.  In particular, the phase of the $\ket{1}\leftrightarrow\ket{r}$ coupling is sinusoidally modulated as defined by a parameterized  time-optimal gate protocol.  The $\sf CZ$ gate parameters (detuning, duration, phase modulation rate, phase modulation depth, phase modulation delay) are optimized in a noiseless simulation which includes the physical error mechanisms such as Rydberg state decay and intermediate state scattering, as well as gate infidelity due to $\sf CZ$ parameters under finite blockade, but no shot-to-shot fluctuations.  Then shot-to-shot noise is applied to each of 10000 samples from distributions of the atom position, atom velocity, laser intensity, beam pointing, and calibration tolerances.  2-photon detuning noise is introduced both through Doppler shifts due to sampling of the atom velocity, and also modification to near-resonant and far-detuned Stark shifts \cite{Maller2015} due to both laser intensity fluctuation and sampling of atom position in the beams.  Additionally, a composite of laser frequency noise across timescales is sampled as a shot-to-shot Rydberg detuning noise, uncorrelated between atoms due to using separate laser systems.  Our temperature measurements and ground-state $T_2^*$ measurements are consistent with a magnetic field noise of std. dev. 5 mG, which contributes std. dev. 7~$\rm{kHz}$ of correlated Rydberg detuning noise.  Electric field noise of $\sigma=5~\rm{mV/cm}$ is assumed and also contributes an average of 7~$\rm{kHz}$ correlated Rydberg detuning noise.  This simulation only includes shot-to-shot fluctuations so does not account for spectrally dependent laser noise.  Although velocity is included in that we account for Doppler shift, we do not model any dephasing due to change in atom position during the gate.  We also do not account for polarization errors that could lead to coupling to the Rydberg $m_j=+\frac{1}{2}$ state.

\begin{table}[!t]
    \caption{Parameters used in Monte Carlo simulations of $\sf CZ$ infidelity. a.u. is atomic units. ``..." indicates no change from current values.}
    \centering
    \begin{tabular}{|l|c|c|}
    \hline
        parameter & current value & projected value \\
        \hline
        trap waist & 1.7 $\mu \rm m$ & 1.0 $\mu \rm m$ \\
        atom temperature & 4 $\mu \rm K$ & 2 $\mu \rm K$ \\
        trap power per site & 2.8 $\rm{mW}$ & 40 $\rm{mW}$ \\
        atom separation & 5.85 $\mu \rm m$ & 4.2 $\mu \rm m$ \\
        blockade strength & 12 $\rm{MHz}$ & 65 $\rm{MHz}$ \\
        2 photon Rydberg Rabi & 1.2 $\rm{MHz}$ & 2.5 MHz \\
        red fractional pulse energy fluctuation (std. dev.) & 0.02 & .001 \\
        blue fractional pulse energy fluctuation (std. dev.) & 0.02 & .001 \\
        correlated Rydberg detuning from magnetic field noise (std. dev.) & 7 $\rm{kHz}$ & 0.5 $\rm{kHz}$ \\
        correlated Rydberg detuning from electric field noise (std. dev.) & 7 $\rm{kHz}$ & 0.5 $\rm{kHz}$ \\
        uncorrelated Rydberg detuning from laser freq. noise (std. dev.) & 1 $\rm{kHz}$ & \ldots  \\
        radial dynamic pointing error (std. dev.) & 40 $\rm{nm}$ & \ldots \\
        radial static pointing calibration error & 50 $\rm{nm}$ & \ldots \\
        Rydberg Rabi rate calibration mismatch (uniform width) & 0.01 & \ldots \\
        trap wavelength & 1064 $\rm{nm}$ & \ldots \\
        \hline
        Rb Rydberg state  & 63$s_{1/2}, m_j\!=\!-\sfrac{1}{2}$ & \ldots \\
        Rb trap polarizability & 698~ a.u. & \ldots \\
        Rb Rydberg lifetime & 112 $\mu \rm s$ & \ldots \\
        Rb intermediate state detuning & -1.4 $\rm{GHz}$ & -10.0 $\rm{GHz}$ \\
        Rb blue DLS frequency & -1.2 MHz & \ldots \\
        Rb blue single-photon Rabi frequency & 90 $\rm{MHz}$ & 344 $\rm{MHz}$ \\
        Rb red single-photon Rabi frequency & 38 $\rm{MHz}$ & 146 $\rm{MHz}$ \\
        Rb red/blue single-photon Rabi frequency ratio & 0.42 & 0.42 \\
        Rb blue waist & 3.45 $\mu \rm m$ & \ldots \\
        Rb red waist & 7.28 $\mu \rm m$ & \ldots \\
        \hline
        Cs Rydberg state & 65$s_{1/2}, m_j\!=\!-\sfrac{1}{2}$  & \ldots \\
        Cs trap polarizability & 1153~ a.u. & \ldots \\
        Cs Rydberg lifetime & 115 $\mu \rm s$ & \ldots \\
        Cs intermediate state detuning & -1.4 $\rm{GHz}$ & -10.0 $\rm{GHz}$ \\
        Cs blue DLS frequency & -1.2 $\rm{MHz}$ & \ldots \\   
        Cs blue single-photon Rabi frequency & 88 $\rm{MHz}$ & 317 $\rm{MHz}$ \\
        Cs red single-photon Rabi frequency & 38 $\rm{MHz}$ & 158 $\rm{MHz}$ \\
        Cs red/blue single-photon Rabi frequency ratio & 0.43 & 0.50 \\
        Cs blue waist & 4.48 $\mu \rm m$ & \ldots \\
        Cs red waist & 3.71 $\mu \rm m$ & \ldots \\
        \hline
    \end{tabular}
    \label{tab:parametertable}
\end{table}

The parameters for the simulation are given in Table~\ref{tab:parametertable}.  The {\sf CZ} baseline error from this simulation with all mechanisms, at our current experimental parameters, is $1-{\mathcal F}=0.0232(2)$  which is close but slightly better than the experimental result, as expected given the unmodeled and unknown error sources.  The calculated error is based on simultaneous Monte-Carlo sampling of all error channels. The result obtained in this way agrees well with a linear sum of the errors from each channel calculated separately. 

In Table \ref{tab:errortable} we report estimated error contributions evaluated by excluding each particular error mechanism with all other mechanisms still present in the simulation.  As such the total simulated error is not a simple summation of the individual mechanisms, but rather each entry represents the potential error reduction available from addressing that mechanism in isolation.  Results given are calculated by removing each mechanism from both atoms and so represent the sum of the effects on both species.

\begin{figure}[!t]
    \centering
    \includegraphics[width=.95\columnwidth]{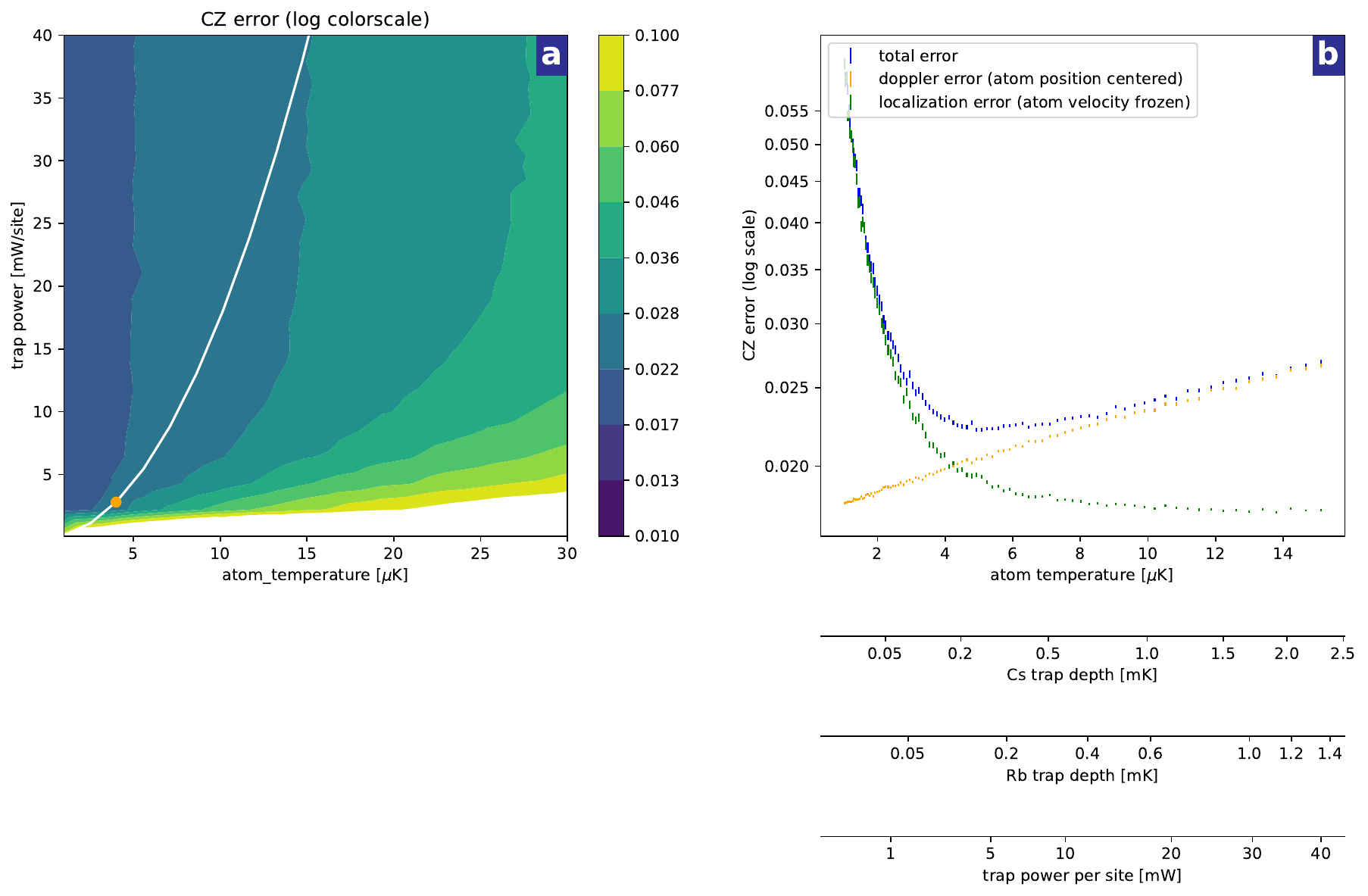}
    \caption{Numerical simulations of gate error. 
        \textbf{a)} A 2D simulation of {\sf CZ} error vs. atom temperature and trap power.  The white line represents the $T\sim \sqrt{U}$ adiabatic expansion cooling curve, and the orange dot represents experimental test conditions at 4 $\mu \rm K$ and 2.8 mW/site trap power. The color scale is $\log_{10}\text{{\sf CZ} error}$.
        \textbf{b)} A trace of predicted {\sf CZ} error through the adiabatic cooling curve.  In experiment, we start at 40 mW/site trap power and lower the trap adiabatically down to 2.8 mW/site and observe the expected cooling of the atoms as $T\sim\sqrt{U}$.  This figure shows simulation results along that constrained 1D trace of trap power versus atom temperature.   As we lower the traps, colder atoms reduce doppler errors, but reduced localization increases positioning errors (including blockade fluctuation).  The $y$-axis is CZ error in linear units, with log spacing, and marker height indicates the $1\sigma$ error bar of the Monte Carlo results.
    }    
    \label{fig:CZ_vs_atom_temp}
\end{figure}

The error mechanisms are isolated in the model in various ways:
\begin{enumerate}
\item
Intermediate state decay: The model Hamiltonian is a three-level system ($\ket{0}, \ket{1}, \ket{r}$) that does not explicitly include the intermediate state, but includes decay of $\ket{1}$ and $\ket{r}$ due to off-resonant coupling with the intermediate state, and that coupling is assumed to produce loss out of the three-level system.  This contribution is isolated by comparison to a system that does not include those decay terms.
\item
Rydberg state decay: The model Hamiltonian includes decay out of the Rydberg state due to the finite Rydberg state lifetime, which is assumed to produce loss out of the three-level system.  This contribution is isolated by comparison to a system that does not include this decay term.
\item
Atom velocity and localization:  The model samples from the atom velocity and position distributions that are calculated for the atom temperature and trap potential.  We can independently zero either the velocity or position, and so we isolate these error contributions by comparing to a system that has one or the other effect removed.  The atom velocity is used for Doppler shift calculations and affects the Rydberg detuning.  The localization term will include the effects of blockade fluctuation, and also finite beam size that impacts lights shifts and Rydberg detuning.  The blockade fluctuation is the dominant localization effect for our beam sizes.  The sum of localization effects can be isolated by freezing the atom position, but as described below we separately isolate the blockade and finite beam size error contributions.  Removing the localization contribution will still allow for finite beam size effects due to static misalignment.
\item
Blockade fluctuation:  The model uses a nominal blockade strength which we get from measurements (see Fig. \ref{fig:rydberg_rabi_data}d in the main text) or from calculation (see Fig. \ref{fig:blockade-in-tweezer}a).  This blockade is then modified shot-to-shot as $r^{-6}$ using the sampled atom positions, and this blockade fluctuation can be isolated by freezing it independently of other atom position effects.
\item
Finite beam size:  This is used to model the effect of atom position and beam misalignment, by calculating the intensity of the Gaussian beams at the atom.  This error contribution is minimal in our baseline system, but its effect can be seen in Fig. \ref{fig:CZ_vs_atom_temp}b as the atom localization becomes poor compared to the beam size at the left-hand side of the plot.  We isolate this contribution by comparing to an infinitely large beam with the same on-center intensity.
\item
Pulse energy fluctuations:  These fluctuations are sampled shot-to-shot for each laser, and are used to modify the light shifts and therefore the Rydberg detuning, as well as the Rydberg Rabi frequencies.  This contribution is isolated by comparison to a system with no fluctuations.
\item
Magnetic and electric field noise:  These contributions as discussed above are both implemented as a correlated Rydberg detuning noise, and can be isolated by removing either $\sigma=7~\rm{kHz}$ contribution from the total correlated noise contribution.
\item
Laser frequency noise:  This is modeled as an additional 2-photon detuning, with uncorrelated samples for each atom.  In simulation this error contribution can be isolated independently of laser intensity noise, or atom position and velocity effects that also contributed to the 2-photon Rydberg detuning.  This noise is not modeled with any spectral dependence, and the value is chosen by comparison between a system that includes Doppler effects, laser frequency noise, magnetic, and electric field noise, versus a system that instead uses total shot-to-shot Rydberg detuning fluctuations consistent with the measured ground-Rydberg decoherence time.
\item
Beam pointing dynamic fluctuation and static misalignment:  Beam pointing fluctuation represents the shot-to-shot AOD aiming and system vibration noise that affects the laser intensity at the atoms, and therefore Rydberg Rabi rate and Rydberg detuning.  Beam misalignment represents the static alignment error that results from an imperfect calibration run, and therefore has any overall intensity error removed in later calibration steps, but still affects the stability due to further movement in the beam because the gradients are bigger off-center.  The effects can be individually silenced.
\item
Rydberg Rabi mismatch:  This represents error where our initial calibration steps do not hit the precise target for laser intensity tuning.  A uniform distribution will be used due to the nature of the calibration algorithm that halts within a certain tolerance.  The overall level will have minimal impact because later calibration of the $\sf CZ$ gate parameters can account for a slight variation in the Rydberg Rabi frequency by modifying the gate duration and other parameters.  However the individual beam intensities are not changed during gate optimization, and so a mismatch between the Rydberg Rabi rates will affect gate performance.
\end{enumerate}

One particularly revealing dissection of error in our system is the balance of localization versus temperature that we achieve from the adiabatic cooling ramp.  During this process we ramp down the trap power, and cooling follows nearly $T\sim\sqrt{U}$.  This follows the white curve in Fig. \ref{fig:CZ_vs_atom_temp}a which show the path that adiabatic cooling takes through the {\sf CZ} landscape.  We make this more clear in Fig. \ref{fig:CZ_vs_atom_temp}b which shows {\sf CZ} error along that constrained 1D path, with the $x$-axis linear in atom temperature, and with {\sf CZ} error on a log scale on the $y$-axis.  Furthermore in Fig. \ref{fig:CZ_vs_atom_temp}b we break down the error contributions.  In one curve we force the atom position to always be centered to zero to show mainly the Doppler contribution to error.  While in the other curve we force the atom velocity to always be frozen to zero to show mainly the localization contribution to error.  As trap power and temperature decrease, the Doppler error is reduced, but localization error increases.  These two competing effects yield an optimal operating point near our experimental best settings.

\begin{table}[!t]
    \caption{Contributions to $\sf CZ$ infidelity from different error mechanisms. The center column are errors found from current parameters and the last column gives projected errors with improvements in Table \ref{tab:parametertable}.  Contributions are calculated as the difference between the baseline system, and the baseline system with all but one mechanism present.}
    \centering
    \begin{tabular}{|l|l|l|}
        \hline
        mechanism & error contribution & projected \\
        \hline
        intermediate state decay & 0.0069(2) & 0.00095(1) \\
        pulse energy fluctuation (blue) & 0.0045(2) & 0.00002(1) \\
        Rydberg state decay & 0.0032(3) & 0.00165(1) \\
        Doppler (atom velocity) & 0.0030(2) & 0.00032(1) \\
        blockade fluctuation (localization) & 0.0026(2) & 0.00001(1) \\
        pulse energy fluctuation (red) & 0.0021(3) & 0.00031(1) \\
        magnetic field noise & 0.0007(2) & 0.00001(1) \\
        electric field noise & 0.0007(2) & 0.00001(1) \\
        finite beam size - blue (localization) & 0.0006(2) & 0.00010(1) \\
        finite beam size - red (localization) & 0.0003(2) & 0.00025(1) \\
        Rydberg Rabi mismatch & 0.0003(2) & 0.00007(1) \\
        laser frequency noise & 0.0002(2) & 0.00003(1) \\
        beam misalignment static & 0.0000(2) & 0.00017(1) \\
        beam pointing fluctuation & 0.0000(2) & 0.00009(1) \\
        \hline
        total error & 0.0232(2) & 0.003403(9) \\
        linear sum ~ $\sum_i \epsilon_i$ & 0.0251(9) & 0.00400(5) \\
        quadrature sum~$\sqrt{\sum_i \epsilon_i^2}$ & 0.0100(2) & 0.00199(1) \\
        \hline
    \end{tabular}
    \label{tab:errortable}
\end{table}

We then make a projection for future performance based on several system improvements.  See Table~\ref{tab:parametertable} for a comparison of the changes.  We predict that Raman sideband cooling will yield an atom temperature of $2~\mu\rm K$.  Furthermore it will allow us to cool efficiently without lowering of the traps during expansion cooling, yielding traps that use the full 40~$\rm{mW}$ available per site.  An improved SLM subsystem is predicted to give trap waists of 1~$\mu \rm m$.  Furthermore, it will allow closer spacing of the traps  for which we predict we will be able to reduce the diagonal Cs-Cs distance to a conservative 6~$\mu \rm m$ which is the current working distance in our best single-species work \cite{Radnaev2025} and thereby give a Rb-Cs distance of 4.2~$\mu \rm m$. This takes advantage of the lack of cross-talk between species, an inherent feature of a dual species system.   At this distance the predicted blockade strength will increase to 65~$\rm{MHz}$.  We also assume improvements to magnetic and electric field noise.  Finally, we will increase the intermediate state detuning to -10~$\rm{GHz}$ and increase the ground-Rydberg Rabi rate to 2.5 MHz on both species.  All other parameters such as Rydberg beam sizes and Rydberg levels are kept the same as the current system.  The specifications will require increases to laser intensity at the atoms
(421~$\rm{nm}: 14\times$,
459~$\rm{nm}: 13\times$,
1005~$\rm{nm}: 15\times$,
1040~$\rm{nm}: 17\times$) through either reduction of beam waists, amplification, or reduced loss in the optical train.  These improvements yield a predicted $\sf CZ$ error of $1-{\mathcal F}=0.003403(9)$.  Further improvements beyond this are possible, here we present only relatively straightforward next steps.  Even larger intermediate state detuning, higher 2-photon Rabi frequency, and longer Rydberg lifetime at higher Rydberg $n$ or colder atom temperatures are all possible directions for further improvement.

\section{QND Measurements}

\begin{figure}[!t]
     \centering
     \includegraphics[width=0.7\textwidth]{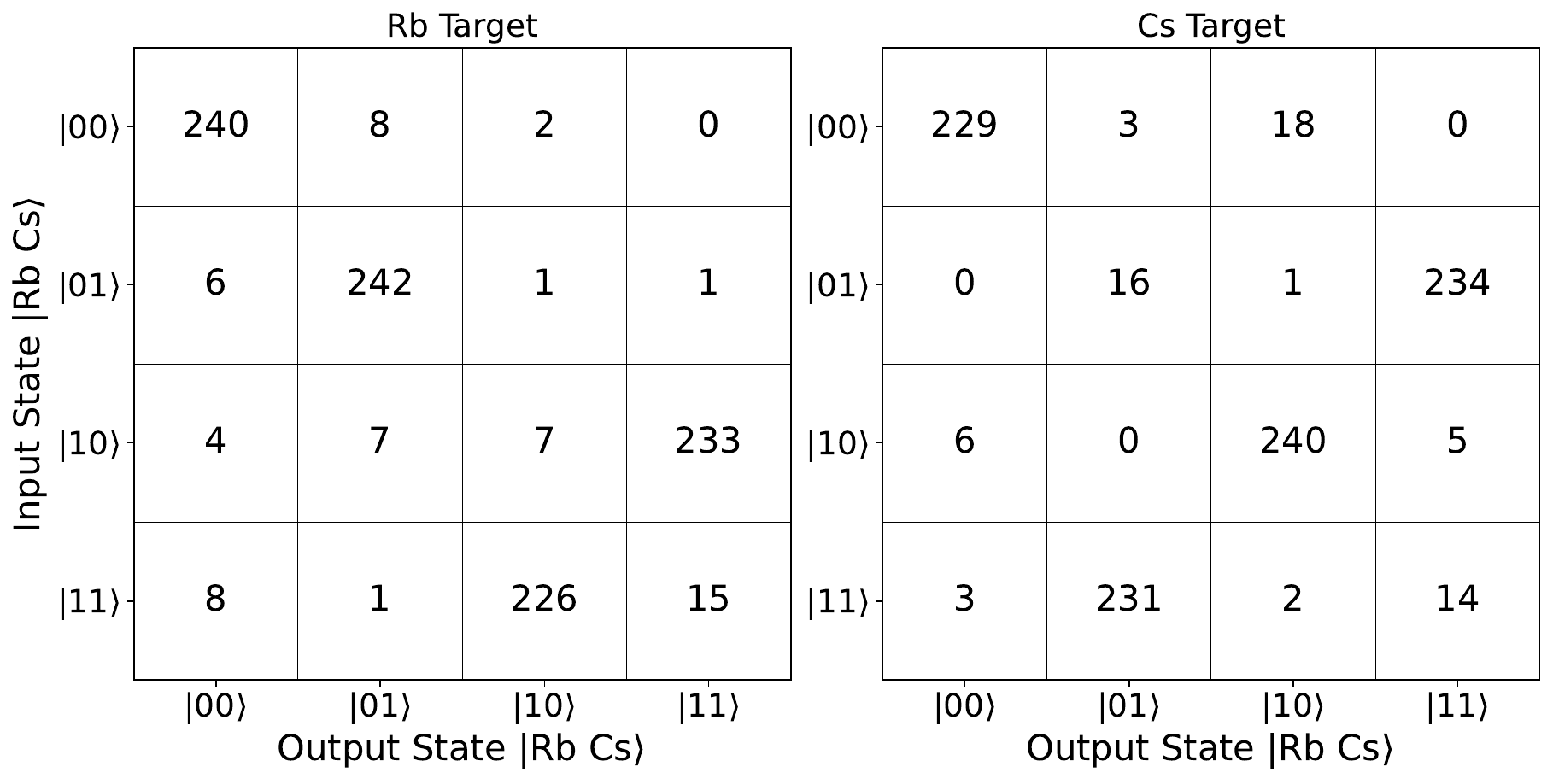}
     \caption{Raw measurement results for two-atom QND.}  
     \label{fig:QND_2atom_numbers_raw}
 \end{figure}

\begin{figure}[!t]
     \centering
     \includegraphics[width=0.7\textwidth]{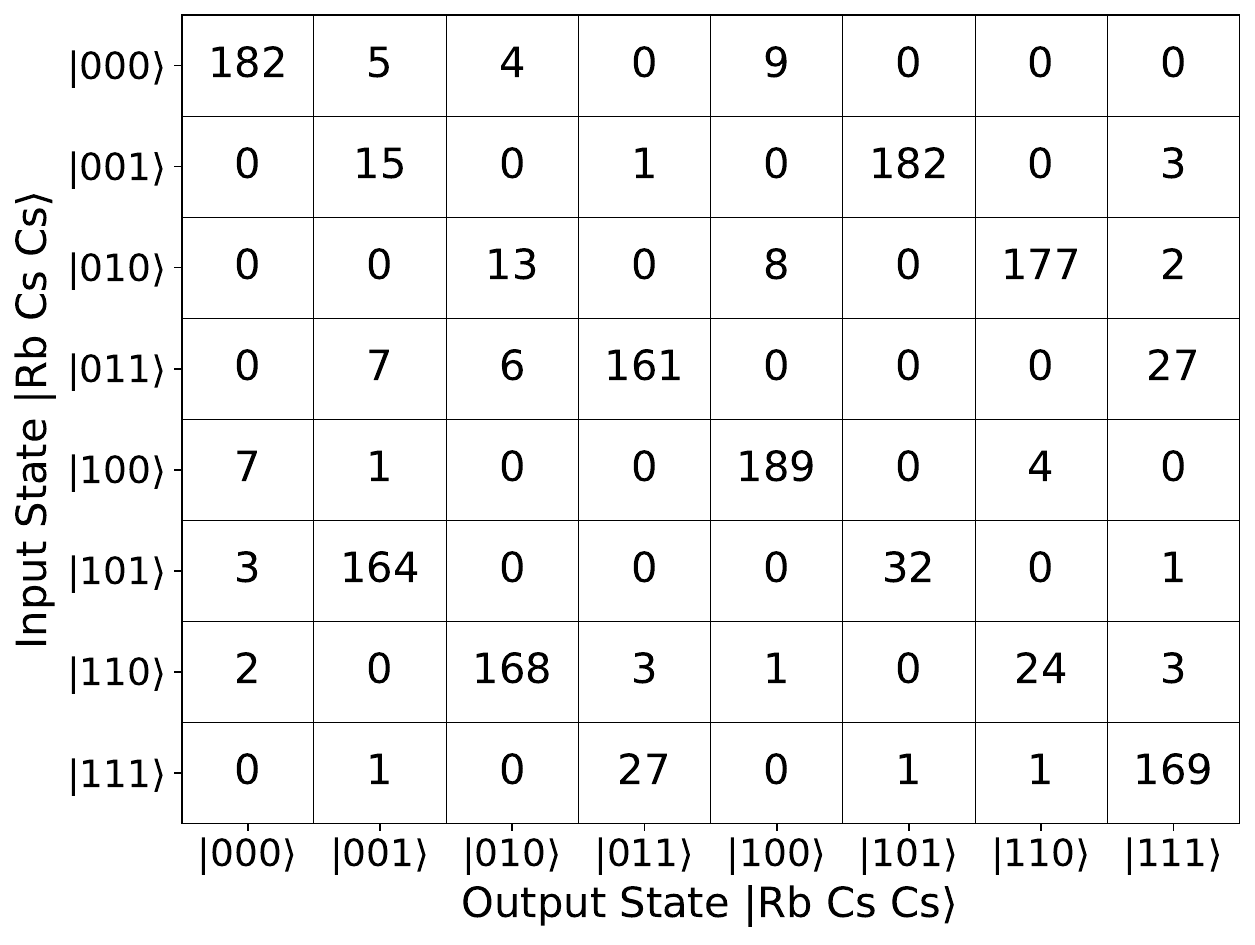}
     \caption{Raw measurement results for three-atom QND.}  
     \label{fig:QND_3atom_numbers_raw}
 \end{figure}

We use Dirichlet statistics to model measurement distributions, binning our data into two categories corresponding to either measuring the correct or the incorrect results.  For each input state for the 2-atom QND measurements, Dirichlet statistics calculated the success probability by comparing the expected output state and the sum of the three incorrect output states.  This success probability was determined for each input state, and then an average of these success probabilities is reported as $\mathcal{F_{\mathrm{QND}}}$.  In the abstract, the 2-atom $\mathcal{F_{\mathrm{QND}}}$ is presented as the average of the Rb and Cs target QND measurements.
The raw measurement counts are shown in Figs. \ref{fig:QND_2atom_numbers_raw} and \ref{fig:QND_3atom_numbers_raw}.

\end{document}